\documentclass[]{jfm}

\usepackage{graphicx}
\usepackage{newtxtext}
\usepackage{newtxmath}
\usepackage{natbib}
\usepackage{hyperref}
\hypersetup{
    colorlinks = true,
    urlcolor   = blue,
    citecolor  = black
}

\newcommand{\RomanNumeralCaps}[1]
\linenumbers

\title{Rayleigh-Taylor Unstable Flames: Thin and Thick}

\author{E.P. Hicks\aff{1}
  \corresp{\email{ephicks@fastmail.net}}}

\affiliation{\aff{1}CIERA, Northwestern University, Evanston, IL 60201, USA\aff{2}Epsilon Delta Labs, Evanston IL, USA}

\begin{document}
\maketitle

\begin{abstract}
A Rayleigh-Taylor (RT) unstable flame is a thin burning interface sandwiched between heavy fuel and light ash layers. RT unstable flames play an important role in complex systems like novel aviation turbine engines, storage facilities for alternative fuels and refrigerants and Type Ia supernovae. Simulations of these systems must use subgrid models of RT flame behavior, but choosing the subgrid model is difficult because RT unstable flames have characteristics of both the classical RT instability and turbulent combustion. In this paper, we investigate whether the flame structure of RT unstable flames can be described using ideas from turbulent combustion theory. We use a large parameter study of Boussinesq model flames and direct measurements of the internal flame structure to show that RT unstable flames can be thickened by their own self-generated turbulence, but that the structure of these thickened flames differs from turbulent flames. Finally, we discuss the implications for modelling RT unstable flames in practical applications.
\end{abstract}


{\bf MSC Codes }  {\it(Optional)} Please enter your MSC Codes here

\section{Introduction}
\label{sec:introduction}

Two fluid layers are Rayleigh-Taylor (RT) unstable if the heavy fluid is accelerated into the lighter fluid. For example, a layer of oil resting on a layer of water is RT unstable because the oil is more dense than the water. If the interface between the oil and water is perturbed, bubbles of water will rise and spikes of oil will fall as the two fluids attempt to change places. The RT instability appears in many places: the Crab Nebula \citep{hester1996,porth2014}, the solar corona \citep{hillier2018,jenkins2022}, salt domes \citep{zaleski1992}, Earth's ionosphere \citep{shinagawa2018}, and in inertial confinement fusion \citep{takabe1991, betti2006, casner2015, martinez2015, yan2016, zhang2018, zhang2018a, xin2019, remington2019}. The classical RT instability is very well studied \citep{sharp1984, bychkov2007, andrews2010, anisimov2013, livescu2013, boffetta2017, zhou2017a, zhou2017,zhou2019, banerjee2020, livescu2020, schilling2020, zhou2021}.

Some important applications include an additional physical effect: burning at the interface between the light and heavy fluids. If the burning layer is thin, it is called a `flame'. RT unstable flames are interesting because they occupy a middle ground between the classical RT instability and turbulent combustion. In this paper, we will explore the flame structure of RT unstable flames and how it is influenced by the competition between the RT instability and the turbulence generated by the flame itself.

\subsection{Applications}

In this subsection, we introduce three applications for which simple models of RT unstable flames are needed.

\subsubsection{Aviation Engines}

One application of RT unstable flames research is to the development of lighter weight, more efficient aviation gas turbine engines. Some engine designs, like the high-g cavity ultra compact combustor inter-turbine burner, rely on using the centrifugal force to drive the RT instability, increasing the rate of fuel consumption so the engine can be smaller \citep{lapsa2009,briones2015}. These complex engine designs are difficult to study, so aviation engineers have developed simpler systems to evaluate the effects of the RT instability on the burning rate. These test systems include both experiments \citep{lewis1971, lewis1973, lewis1975,lapsa2009,erdmann2019, erdmann2023} and simulations \citep{briones2015, sykes2019, sykes2021}. Some simulations rely on flamelet models to simulate combustion. For example, \citet{sykes2019, sykes2021} used a compressible flamelet model in their 3D LES of burning in a curved duct. Using flamelets requires making assumptions about how the flame responds to turbulence. Sykes chose to use no efficiency function because \citet{hicks2015,hicks2019} showed that the RT instability thins, rather than thickens, the flame. In this paper, we will show that the flame does thicken in certain regimes and make recommendations for flamelet model choices.

\subsubsection{Alternative Fuels and Refrigerants}

Another terrestrial application of RT unstable flames research is to alternative fuels and refrigerants with lower global warming potential. For example, ammonia burns without emitting CO2 and can be stored and transported more easily than hydrogen, although care must be taken to prevent ammonia leaks and the emission of N2O (a potent greenhouse gas) and NOx (which cause smog and acid rain) \citep{bertagni2023}. The newer refrigerants R-32 and R-1234yf have lower global warming potential than older refrigerants and do not deplete ozone \citep{mclinden2017}.

Ammonia, R-32 and R-1234yf are all mildly flammable when mixed with air, so care must be taken with their transport, storage and use. In particular, R-32 and R-1234yf replace refrigerants that are not flammable, potentially creating new hazards. As a result, there has been an effort to quantify the fire and explosion risk of these fuels and refrigerants \citep{davis2017, pagliaro2017, papas2019, hesse2021, hegetschweiler2022, burgess2022, domanski2023, hesse2023, tavares2023, glaznev2024, mathew2025}.

One aspect of quantifying risk is understanding the effect of buoyancy on these flames. These mildly flammable mixtures have especially low laminar flame speeds, which makes them more susceptible to buoyancy effects than faster flames like methane/air or hydrogen/air. For example, \citet{davis2017, pagliaro2017} describe very large experiments in which ammonia flames are clearly seen to rise as buoyant bubbles. \citet{papas2019} describe a buoyant bubble of burning R-1234yf/air with ``jet-like flame features'' on its surface. These flames are also susceptible to the Rayleigh-Taylor instability. \citet{tavares2024} recently measured linear growth rates for R-32/air flames using DNS and showed that the RT instability can enhance the flame speed in the nonlinear phase. Understanding exactly how much the RT instability can enhance the flame speed in large storage containers is key to using these new fuels and refrigerants safely. Simulations of large domains, which are typically RANS or LES, rely on flamelet-based combustion models which are usually based on the physics of turbulent flames. Therefore, it is important to determine whether RT unstable flames actually behave like turbulent flames or whether different models are needed when buoyancy is important.

\subsubsection{Type Ia Supernovae}

An important astrophysical application of RT unstable flames research is to the study of Type Ia supernovae. Type Ia supernovae are extremely bright stellar explosions which were used to show that the universe is accelerating in Nobel-prize winning research by \citet{riess1998,perlmutter1999}. Given the importance of Type Ia supernovae, astrophysicists have put a great deal of effort into attempting to determine what kinds of stellar systems produce Type Ia supernovae and how the explosion happens. This is difficult because there are many competing progenitor models \citep{ruiter2025}. Even worse, no single progenitor model can explain all of the observational data and it is now widely accepted that Type Ia supernovae are produced through multiple channels. The only way to determine which progenitor systems may have produced any given supernova is to compare its light curve, nuclear yields, and/or ejecta properties with simulated output from different progenitor candidates \citep{blondin2022,tiwari2022}.

Candidate progenitor light curves and nuclear yields are generated from complex simulations of one or more white dwarf stars \citep{tiwari2022}. These simulations must include models for the thermonuclear burning that drives the explosion. Since it is impossible for simulations to resolve the thin flame, they must use models that describe the flame's behavior \citep{oran2005}. For example, simulations must include a subgrid model for the flame speed \citep{khokhlov1995,niemeyer1995,khokhlov1996,niemeyer1997,niemeyer1997a,reinecke1999,gamezo2003,schmidt2006a,schmidt2006b,zhang2007,jackson2014}. For progenitor systems in which the flame is expected to be RT unstable, there are two sources of turbulence that must be considered: pre-existing turbulence generated by convection \citep{nonaka2012} and the expansion of the white dwarf, and turbulence that is generated by the RT unstable flame itself. One important question is whether the flame is affected by these sources of turbulence in the same way.

In addition, in some progenitor systems, the explosion is thought to be driven by a transition from a subsonic flame (a deflagration) to a supersonic flame (a detonation) \citep{blinnikov1986,woosley1990,khokhlov1991,khokhlov1997,khokhlov1997a,gamezo2004,ropke2007,seitenzahl2013}. Understanding how this deflagration-to-detonation (DDT) transition occurs requires understanding the physics of the flame, in particular how its structure is affected by turbulence. Overall, understanding the fundamental physics of RT unstable flames and whether they can be modeled as turbulent flames is important for choosing Type Ia simulation subgrid models and for understanding how these stellar systems explode.

\subsection{Our Goal: Assessing Simple Models of RT Unstable Flames}

In all three of these applications, there is a need for simple models that describe the behavior of RT unstable flames. If RT unstable flames could be shown to behave like some other form of combustion for which simple models have already been developed, then those models could be adopted by analogy. The most promising option is turbulent combustion, but RT unstable flames and turbulent combustion differ in key ways. The field of turbulent combustion considers a flame consuming turbulent fuel~\citep{peters2000,driscoll2020}. The flame is forced to interact with turbulence in order to propagate and the flame's properties are shaped by a two-way competition between turbulence and burning~\citep{poludnenko2010,poludnenko2011a,hamlington2011,hamlington2012}. On the other hand, RT unstable flames generate their own turbulence. When the RT instability is strong, it stretches the flame front into bubbles and spikes, substantially increasing the flame's surface area and flame speed~\citep{khokhlov1995}. The sides of the bubbles and spikes generate turbulence baroclinically~\citep{V03}. This self-generated turbulence is washed downstream as the flame propagates upwards and the front layers of the flame may not interact much with it. Meanwhile, burning works to destroy the surface area created by the RT instability. So, the properties of RT unstable flames are determined by a three-way competition between the RT instability, self-generated turbulence and burning~\citep{hicks2015,hicks2019}. Here, we will explore how the structure of RT unstable flames responds to self-generated turbulence and whether this response can be understood using ideas from turbulent combustion theory.

In this paper, we will consider four questions that shed light on whether RT unstable flames can be modelled using simple models from turbulent combustion. All of these questions focus on whether RT unstable flames are affected by self-generated turbulence in the same way as laminar flames are affected by upstream turbulence. First, can RT unstable flames be thickened by their own self-generated turbulence? Second, how does the flame structure change during the transition from thin to thick? Third, do RT unstable flames follow the turbulent combustion regime diagram? Fourth, can RT unstable flames transition to distributed burning? Together, these questions establish whether the flame structure of RT unstable flames can be described using ideas from turbulent combustion.

\subsection{Previous Work}

Our first question is whether RT unstable flames can be thickened by their own self-generated turbulence. Turbulent flames are thickened by turbulence if it is strong enough. We found in our previous work \citep{hicks2015,hicks2019} that the RT unstable flames in our large parameter study (which varied gravity and the box size) were typically thinner than laminar. It appeared that stretching by the flame front by the RT instability thinned the flame and overwhelmed any thickening due to self-generated turbulence. We concluded that RT unstable flames are `thin' and dissimilar to turbulent combustion. However, our work considered flames confined by their computational domains. In contrast, \citet{chertkov2009} and \citet{ley2024} considered the evolution of unconfined reactive RT mixing zones. \citet{chertkov2009} investigated how the thickness of the interface between fuel and ash evolves with time for the case where the reaction is much slower than the RT instability. They used phenomenological arguments to predict that the interface between fuel and ash in an unconfined, self-similar growing reactive mixing layer will become thicker with time even as the flow becomes more segregated and the amount of mixing between fuel and ash decreases. \citet{ley2024} investigated the effects of molecular mixing on the reactive RT mixing zone and showed that higher mixing slows the growth of the mixing zone. These authors infer that that the flame will be thickened by self-generated turbulence as the mixing zone grows, but they did not measure the interior flame structure directly. In this paper, we use direct measurements of the flame structure to reconcile \citet{hicks2015,hicks2019} with \citet{chertkov2009,ley2024}.

Our second question is how the flame structure changes during the transition from thin to thick. Thickening in turbulent flames begins in the preheat (front) layer of the flame while the reaction zone remains laminar in structure. Is this also the way that RT unstable flames thicken? Although many numerical studies have investigated the dynamics of RT unstable flames or reactive RT~\citep{khokhlov1994, khokhlov1995, V03, V05, bell2004b, zingale2005a, zhang2007, chertkov2009, biferale2011, hicks2013, hicks2014, hicks2015, hristov2018, hicks2019, liu2024, ley2024, zheng2025}, none have directly measured the interior structure of the flame itself.  We present measurements of the internal flame structure in this paper.

Our third question, whether RT unstable flames follow the turbulent combustion regime diagram, is also mostly unexplored. Studies that discuss the regime diagram for RT unstable flames typically assume that RT unstable flames behave as turbulent combustion predicts, calculate the relevant non-dimensional numbers (like the Karlovitz and Damkohler numbers) and use those numbers to assign the simulation to the corresponding turbulent combustion regime. In this paper, we will expand on our previous work \citet{hicks2015,hicks2019}, which classified RT unstable flames as `flamelets' based on the fact that they are typically thinner than laminar, and use the internal structure measurements to consider whether RT unstable flames behave like turbulent flames at all. We will show that some ideas from turbulent combustion can be adapted for RT unstable flames, but that a new regime diagram is required.

Our fourth question is whether RT unstable flames can be so disturbed by their own self-generated turbulence that they enter the distributed burning regime. Distributed burning is an `extreme' turbulent combustion regime in which the dynamics of the inner reaction zone of the flame are modified by turbulence \citep{damkohler1940,aspden2019,driscoll2020}. There are a few slightly different definitions of distributed burning, but all agree that the flame will become significantly thicker than laminar \citep{aspden2011,skiba2018,driscoll2020}. Distributed burning is difficult to achieve in terrestrial settings \citep{aspden2019,shi2024}, but may play a key role in the explosions of Type Ia supernovae \citep{aspden2008,woosley2009,aspden2010}. According to one explosion scenario, the flame undergoes the DDT that leads to the supernova explosion because the flame enters the distributed burning regime~\citep{khokhlov1997,khokhlov1997a, niemeyer1997,lisewski2000,dursi2006,woosley2007,woosley2009,seitenzahl2009}. In this case, DDT occurs via the Zel'dovich gradient mechanism~\citep{zeldovich1970} or SWACER \citep{lee1978}. The conditions for a Zel'dovich gradient mechanism DDT may also be established by the conditioning of the fuel layer ahead of the flame front by adiabatic heating~\citep{brooker2025}. Alternatively, the DDT could be triggered by other mechanisms like turbulent self-acceleration \citep{poludnenko2011,poludnenko2015,poludnenko2017_icders,poludnenko2019}.
In this paper, we will explore whether RT unstable flames can undergo a transition to distributed burning due to their own self-generated turbulence.

\subsection{Our Strategy and Paper Organization}

To find a transition from thin to thick flames, we set up our computational experiment so that we could roughly control the length of the turbulent cascade on scales smaller than the flame width. Controlling the amount of turbulence within the flame front allows us to test the hypothesis, from turbulent combustion, that small eddies thicken the flame. We control the length of the turbulent cascade within the flame front in four ways: setting the gravity ($G$), setting the turbulent integral scale by choosing the width of the computational domain ($L$), controlling the size of the Kolmogorov scale by setting the Prandtl number ($Pr$), and controlling the laminar flame width by changing the reaction model (set by the parameter $p$).

We investigated the effect of varying $L$ and $G$ in our previous work, so in this paper we vary $Pr$ and $p$. Neither the effect of $Pr$ nor the effect of the reaction type has been significantly explored in previous work on RT unstable flames. \citet{V03} showed that the critical onset gravity ($G_{cr}$) at which the Rayleigh-Taylor instability first deforms the flame front and the length of velocity variation behind the flame both depend on $Pr$. Typically, investigators have simply set $Pr=1$ or relied on numerical or subgrid model dissipation schemes without reporting a value for $Pr$. However, it is important to establish the effect of varying $Pr$ because Type Ia supernovae have a very low $Pr$ of $10^{-5}$ \citep{timmes1992}. $Pr$ is much closer to $1$ in terrestrial applications.

Likewise, the effect of varying the reaction model on the dynamics of RT unstable flames hasn't received much attention, both because it is generally assumed that dynamics are driven at large scales and because there was no clear method for systematically varying the reaction model. Simulations of specific reactions, like hydrogen burning or carbon burning in Type Ia supernovae simply use the appropriate chemistry or a reaction model based on the specific reaction (e.g. CF88 model for carbon burning
\citep{caughlan1988}). Studies attempting to draw general conclusions typically select a simple one-step reaction model (for a review of model reaction types see \citep{xin2000}). The most popular choice is the Kolmogorov-Petrovkii-Piskunov (KPP) model, since it produces a easily resolved thick interface, but other choices include a bistable reaction or a `top-hat' reaction with a constant reaction rate and a specified ignition temperature. In this paper, we develop a family of traveling wave flame front solutions with the same laminar flame speed, but different laminar flame widths. We use these solutions to systematically investigate the effect of the reaction model on RT flame dynamics, specifically the effect of self-generated turbulence of the average flame width and on the internal flame structure, which has never been explored.

The paper is organized as follows. We begin by describing our simulations and numerical methods in section \ref{section:methods}. We continue by describing the combustion regimes predicted by turbulent combustion theory in section \ref{section:theory} and how the understanding of these regimes has been changed by recent experiments and simulations in section \ref{section:reality}. Then, in section \ref{section:comparison_machinery} we explain how we will compare RT unstable flames with turbulent flames and formulate the necessary nondimensional numbers. We present our results and answer our four questions in section \ref{section:results}. Finally, we conclude in section \ref{section:conclusions} and discuss the implications of our results in \ref{section:discussion}.

\section{Simulations and Numerical Methods}
\label{section:methods}

\subsection {Equations and Control Parameters}

For simplicity, we simulate the Boussinesq equations with a model reaction term. The Boussinesq approximation assumes that the density jump across the flame front is small and only takes it into account in the forcing term~\citep{spiegel1960}. This allows us to isolate and exclude unwanted phenomena like the Landau-Darrieus instability, shocks, and heating due to viscous dissipation.

We simulate the flame using a simple model reaction with dimensional flame speed $\tilde s_0 = \sqrt{\alpha \kappa}$ and dimensional laminar flame width $\tilde\delta=\sqrt{\kappa / \alpha}$, where $\alpha$ is the laminar reaction rate, and $\kappa$ is the thermal diffusivity. Non-dimensionalizing the Boussinesq equations using $\tilde\delta$ and the reaction time ($1/\alpha$) gives
\begin{subequations}
\begin{gather}
 \dfrac{D \textbf{u}}{Dt}=-\left(\dfrac{1}{\rho_o}\right)\nabla p + G\,T + Pr
\nabla^{2} \textbf{u} \label{NNS}\\
\nabla \cdot \textbf{u}=0  \label{Nincomp}  \\
\dfrac{D T}{D t} = \nabla^{2} T + R(T), \label{Nheat}
 \end{gather}
\end{subequations}
where temperature ($T$) is a reaction progress variable that tracks the fluid from unburnt fuel at $T=0$ to burnt ashes at $T=1$~\citep{V03}. $R(T)$ is the model reaction rate. There are three control parameters:
\begin{align}
G &=g\left(\frac{\Delta \rho}{\rho_o}\right)\dfrac{\tilde\delta}{\tilde s_o^{2}} \\
Pr &=\frac{\nu}{\kappa} \\
L &=\dfrac{\tilde \ell}{\tilde \delta}
\end{align}
where $G$ is the non-dimensionalized gravity, $Pr$ is the Prandtl number, and $L$ is the non-dimensional domain width. $\tilde \ell$ is the dimensional length in the $x$ and $z$ directions. The nondimensional laminar flame speed is $s_0=1$ and the nondimensional laminar flame width is $\delta=1$.

\subsection {p-flames: A Parameterized Family of Traveling Wave Solutions}

In order to consider the effect of reaction steepness, we construct a single-parameter family of laminar flame solutions with different physical flame front widths but with the same laminar flame speed, $s_0=1$.

\begin{figure}
\captionsetup{singlelinecheck = false, justification=justified}
  \centerline{\includegraphics{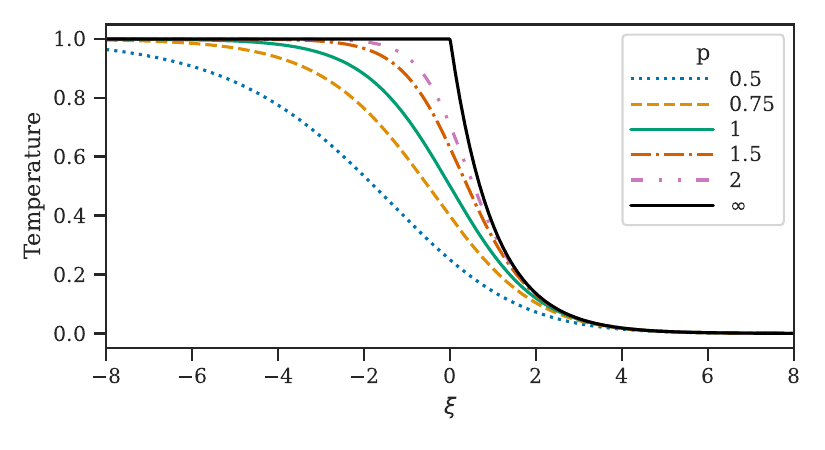}}
  \caption{p-flame temperature profiles for the values of $p$ used in this study and the limiting profile as $p \rightarrow \infty$ (see Appendix \ref{appendix:plimits}).}
\label{figure:pflames}
\end{figure}

In Appendix \ref{appendix:psolution}, we show that the reaction term
\begin{equation}
R(T)=(p+1) T^{p+1}\left(1-T^p\right)
\end{equation}
(where $p>0$ is an adjustable parameter), generates a family of solutions with the explicit traveling wave solution

\begin{equation}
\label{equation:Tprofile}
 T(\xi, p)=\left[\frac{1}{2}-\frac{1}{2} \tanh \left(\frac{p \xi}{2}\right)\right]^{1 / p}
\end{equation}
where $\xi=y-s_0 t$. Figure \ref{figure:pflames} shows this temperature profile for the values of $p$ used in this study. Flames with larger $p$ are thinner and flames with smaller $p$ are thicker.

There are two common definitions for the physical flame width. The thermal flame width is defined as $\delta_T = (T_{\mathrm{max}} - T_{\mathrm{min}}) / \mathrm{max}(|\nabla T|) = 1/ \mathrm{max}(|\nabla T|)$. Using the derivatives for laminar p-flames derived in Appendix \ref{appendix:pderivatives} this becomes $\delta_T(p)=(1/p)(p+1)^{1+1/p}$. In Appendix \ref{appendix:plimits}, we show that $\delta_T(p)\rightarrow1$ as $p\rightarrow\infty$. Alternatively, the contour flame width can be defined as the distance between two temperature contours, $\delta^{T_2}_{T_1}$; in this paper we choose $T_1=0.1$ and $T_2=0.9$. For p-flames, the laminar contour flame width can be found by solving Equation \ref{equation:Tprofile} for $\xi$ and calculating $\delta^{0.9}_{0.1}(p)=\xi(0.1)-\xi(0.9)$. In Appendix \ref{appendix:plimits}, we show that  $\delta^{0.9}_{0.1}(p)\rightarrow2.197...$ as $p\rightarrow\infty$. Laminar values $\delta_T(p)$ and $\delta^{0.9}_{0.1}(p)$ for the values of $p$ used in this paper ($p=0.5,0.75,1,1.5,2$) are given in table \ref{table:flamewidths}. In this study, we compare measurements of $\delta^{0.9}_{0.1}$ of simulated RT unstable flames to their laminar values $\delta^{0.9}_{0.1}(p)$.

\begin{table}
\captionsetup{singlelinecheck = false, justification=justified}
  \begin{center}
\def~{\hphantom{0}}
  \begin{tabular}{lcc}
      $p$  & $\delta_T(p)$   &   $\delta^{0.9}_{0.1}(p)$  \\[3pt]
       0.5   & 6.750 & 7.376 \\
       0.75   & 4.921 & 5.373 \\
       1  & 4.000 & 4.394\\
       1.5   & 3.070 & 3.458  \\
       2 & 2.598 & 3.023  \\
  \end{tabular}
  \caption{Laminar thermal flame width $\delta_T(p)$  and laminar contour flame width $\delta^{0.9}_{0.1}(p)$ for the values of $p$ used in this study. Values are rounded to three decimal places.}
  \label{table:flamewidths}
  \end{center}
\end{table}

\subsection{Simulations}

Simulations were set up in the following way. All simulations were in 3D with the flame propagating upwards in the +y-direction against gravity in the -y-direction. The domain was a long, square shaft with equal length in the x and z-directions and a much larger height in the y-direction. Boundary conditions were periodic on the sidewalls, inflow at the top of the box, and outflow at the bottom of the box. The initial velocity was zero in the entire domain. The top boundary was held at $T=0$ (fuel) and the bottom boundary at $T=1$ (ash). The flame does not approach either the top or bottom boundary during the simulations. The initial flame front was a perturbed plane with a tanh-shaped profile matching the laminar flame profile given in equation \ref{equation:Tprofile}. More details about our general setup are given in \citet{hicks2015, hicks2019}. All simulations were run using Nek5000, a freely available, open-source, highly scalable spectral element code developed at Argonne National Laboratory (ANL)~\citep{nek5000}.

All simulations have $L=32$, $G=32$. We choose this parameter combination because it showed a significant deviation from the RT flame speed model in our earlier study \citep{hicks2015}, but the required box size is small enough to carry out a large sweep in the $(p, Pr)$ parameter space. Generally, we focused on lower $Pr$ simulations since these are most relevant for our applications, but we carried out a few high $Pr$ simulations for comparison. In total, we simulated 24 different $(p, Pr)$ parameter combinations. For $Pr=0.5, 0.75, 1, 2$, we simulated all 5 values of $p$. For $Pr=4, 8$ we simulated $p=0.5, 1$.

Each simulation begins with an initial transient during which larger wavelength modes are generated by the RT inverse cascade~\citep{liu2024}. The flame speed grows as more surface area is produced. After transient growth is complete and the instability has saturated, the flame speed (and other quantities, like the flame width) varies around a statistically steady average. Simulations were run for long times, so that time-averages could be computed using many periods of each oscillating quantity. We ignored the initial transient.

We carried out 2, 3 or 4 simulations at different resolutions for each parameter combination. At least one simulation for each parameter combination has a measured viscous scale smaller than the average resolution of the simulation, thus qualifying as DNS. Other simulations have viscous scales slightly larger than the average resolution, making them nearly DNS. Throughout the paper, we typically show all 64 simulations in the study so that the reader may see the measurement spread for each $(p, Pr)$ parameter combination. Measurement spread may be due to different simulation resolutions, the uncertainty associated with averaging over an oscillating function, or intrinsic variability due to different realizations of flame behavior. Since we don't observe clear trends with resolution and we average over long time series, we believe that the measurement spreads for simulations with the same $(p, Pr)$ are mostly due to intrinsic variability. By showing all of the simulations, we hope to make the amount of variability clear and show that our conclusions are general.

\section{Turbulent Combustion}
\label{sec:turbulentcombustion}

In this section, we build a framework for comparing RT unstable flames with turbulent combustion. We briefly review traditional turbulent combustion theory and the theoretical turbulent combustion diagram. Next, we review recent experiments and simulations that suggest that revisions to the theoretical picture are needed. Finally, we explain our strategy for making comparisons between RT unstable flames and turbulent combustion and summarize the necessary nondimensional numbers.

\subsection{Traditional Turbulent Combustion Theory}
\label{section:theory}

\begin{figure}
\captionsetup{singlelinecheck = false, justification=justified}
  \centerline{\includegraphics[width=5.5in]{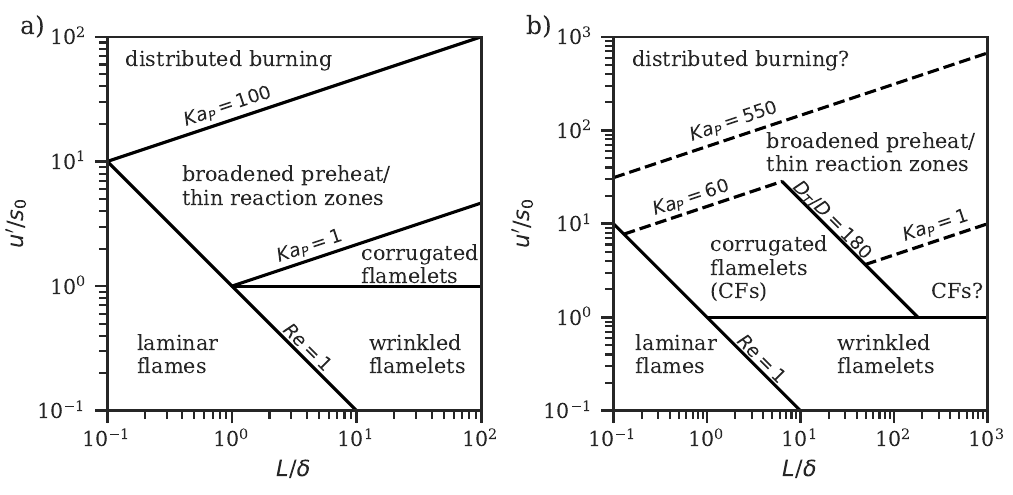}}
\caption{This figure compares the theoretical turbulent combustion regime diagram (panel a, adapted from \protect\citet{peters2000}) to a measured regime diagram based on experiments and DNS (panel b, adapted from \protect\citet{skiba2018}). In the measured regime diagram, boundaries drawn with dashed lines aren't well constrained, but are plausible.}
\label{figure:traditional_experimental}
\end{figure}

One version of the turbulent combustion regime diagram is shown in figure \ref{figure:traditional_experimental}a. This diagram combines diagrams proposed by \citet{borghi1985,peters2000,aspden2010,aspden2019}. A detailed history of how the turbulent combustion regime diagram has evolved in time is given in \citet{aspden2011}. To construct the diagram, ratios of the turbulent velocity, laminar flame speed, integral scale and laminar flame width are combined to make three non-dimensional control parameters. The Reynolds number ($Re$) describes the overall strength of the turbulence relative to dissipation. The Damkohler number ($Da$) describes the strength burning relative to large-scale turbulence. The Karlovitz number ($Ka$) describes the strength of the small-scale turbulence relative to burning. These dimensionless numbers are used to predict changes in the flame behavior based on simple physical reasoning. Predicted turbulent combustion regimes are:

\subsubsection*{\textbf{Laminar Flames ($Re<1$)}}
Turbulence is so weak that it does not affect the flame.

\subsubsection*{\textbf{Wrinkled Flamelets ($Re>1$, $u'/s_o<1$)}}
Even the largest turbulent eddies do not turn over fast enough to compete with burning. The flame structure is laminar.

\subsubsection*{\textbf{Corrugated Flamelets ($u'/s_o<1$, $Ka<1$)}}
All turbulent length scales are larger than the flame width, so the flame structure remains laminar, but turbulence is strong enough to gently wrinkle the flame.

\subsubsection*{\textbf{Broadened Preheat / Thin Reaction Zone ($Re>1$, $Ka>1$)}}
The smallest turbulent eddies are smaller than the flame width and turn over more quickly than the flame burns. These eddies thicken the preheat region of the flame, but the reaction layer remains laminar in structure.

\subsubsection*{\textbf{Distributed Burning ($Ka>100$)}}
Turbulence is strong enough to affect the reaction zone of the flame. No part of the flame has a laminar structure; the entire flame is thickened by turbulence. Propagation of the flame is driven by turbulent diffusion. This regime can be divided into the \textit{\textbf{Stirred Flame ($Da>1$)}} and \textit{\textbf{Well-Stirred Reactor}} ($Da<1$) regimes. If the turbulence is strong enough to extinguish at least $10\%$ of the flame front, then flame is in the \textit{\textbf{Broken Reaction Layers}} regime instead of the Stirred Flame regime \citep{driscoll2020}.

\medskip
\noindent Note that these regimes, their names, and the physical reasoning behind them are not standardized.

\subsection{Turbulent Combustion: Experiments and Simulations}
\label{section:reality}

Over the past twenty years, experiments and simulations have become sophisticated and comprehensive enough to test the theoretical combustion regimes. \citet{driscoll2020} reviews these efforts, focusing on the response of flames to `extreme turbulence'. Significantly, it appears that experimental evidence does not support the theoretical combustion regime diagram discussed in the previous subsection. A new regime diagram proposed by \citet{skiba2018} based on experimental measurements and simulations is shown in figure \ref{figure:traditional_experimental}b.

According to \citet{driscoll2020}, there are a few significant differences between the theoretical diagram and the measured diagram. First, the transition between flames that are simply wrinkled by turbulence (Corrugated Flamelets) and flames with thickened preheat regions (Broadened Preheat, Thin Reaction Layers) is triggered by an increase in the ratio of turbulent diffusivity to material diffusivity instead of an increase in $Ka$. Specifically, experiments find a transition at around $u^\prime L/D \sim 180$. Second, experimental data suggests that the flame only enters the Broken Reactions regime if the flame does not have adequate back support, so that cold air is entrained into the flame and extinguishes part of it. It does not appear that extreme turbulence alone can significantly extinguish the flame. Third, true distributed burning due to extreme turbulence is extremely difficult to achieve and requires a very high $Ka$. Simulations of lean premixed H2/air flames find stirred flames at around $Ka = 1560$ and well-stirred reactors at $Ka = 8700$, much higher than the $Ka = 100$ boundary in the theoretical regime diagram \citet{aspden2011,aspden2019}. Likewise, \citet{shi2024} found distributed flames at $Ka = 7690$ using a lean premixed hydrogen/air turbulent jet flame experiment. Finally, the measured regimes diagram depends on more than the axes of the regime diagram. Fuel type, Lewis number, equivalence ratio and geometry can all alter the regime boundaries. For example, the transition to distributed burning depends on fuel type. \citet{aspden2008,aspden2010} found carbon flames in the distributed burning regime at $Ka$ as low as $230$ but \citet{aspden2019} showed that methane/air flames still aren't in the distributed burning regime at $Ka=8700$. All of these findings suggest that the theoretical turbulent combustion diagram does not accurately describe turbulent combustion.

\subsection{How To Compare Turbulent Combustion with RT Unstable Flames }
\label{section:comparison_machinery}
The current tension between turbulent combustion theory and experiments/simulations makes comparisons with RT unstable flames more difficult. Experiments represent reality, but turbulent combustion theory underpins the modeling of flame behavior in complex applications. To deal with this tension, we will make comparisons with RT unstable flames  that focus on elements of the physical picture shared by both theory and experiments. We will account for differences between the physical pictures by considering a variety of possible physical models and regime transition scenarios. These differences are manifested by the use of different nondimensionless numbers $(Re, Da, Ka)$ to describe the transitions between regimes.

In order to make comparisons between turbulent combustion and our RT unstable flames setup, we need to reformulate the traditional non-dimensional numbers used in turbulent combustion to account for $p$ and $Pr$. In general, we account for $p$ by using a generic, physically appropriate laminar flame width ($\tilde\delta_F(p)$) instead of the dimensional flame width ($\tilde{\delta}$).  In practice, $\tilde{\delta}_F$ could be the thermal flame width $\tilde{\delta}_T(p)$ or the contour flame width $\tilde\delta^{T_2}_{T_1}(p)$ since both quantities represent the physical laminar flame width for a given $p$. In this paper, we use $\tilde{\delta}_{0.1}^{0.9}(p)$. Both choices yield similar qualitative results. We incorporate $Pr$ by retaining $\nu$ in calculations. There is nothing RT-specific about these reformulations, and they could be used in studies of turbulent combustion. An alternative formulation that considers the effect of $Pr$ on the turbulent combustion regimes is given by~\citet{kerstein2001}. We summarize the nondimensional numbers that we use in the next subsection; an extended derivation and discussion can be found in Appendix \ref{appendix:numbers}.

\subsubsection{Summary of Nondimensional Numbers}
\label{section:numbers}

We consider a traditionally defined Reynolds number, which in our nondimensionalization is
\begin{equation}
Re=\frac{u^\prime L}{Pr}.
\label{eqn:Re}
\end{equation}
\noindent
\noindent
To compare with the measured regime diagram of \citet{skiba2018}, we consider the ratio of the turbulent diffusivity and the thermal diffusivity:

\begin{equation}
\frac{D_T}{\kappa} = \frac{\tilde{u^\prime} \tilde{l}}{\kappa} = u^\prime L.
\label{eqn:diff_ratio}
\end{equation}
\noindent
\noindent
We formulate a generic Damkohler number based on the physical flame width:

\begin{equation}
Da=\frac{t_T}{t_F}=\frac{s_0 L}{u^\prime \delta_F(p)}.
\label{eqn:Da}
\end{equation}

We consider three Karlovitz numbers based on the physical assumption that the flame is thickened by eddies that mix the flame faster than it burns. Mathematically, these $Ka$ are based on $t_{F} / t_\eta$ where $t_{F}$ is the flame burning time and $t_\eta$ is the turnover time of the Kolmogorov scale eddies. The first of these numbers is the classical Karlovitz number described in \citet{peters2000}, which we will call $Ka_P$:

\begin{equation}
Ka_P=\left(\frac{u^\prime}{s_0}\right)^{3 / 2}\left(\frac{L}{\delta}\right)^{-1 / 2}.
\label{eqn:Ka_P}
\end{equation}
\noindent
\noindent
This derivation assumes that $\tilde{\delta}= D/\tilde{s_0}$ and $Pr = 1$. The second number ($Ka_x$), which we will call the ``extended Karlovitz number'', is a reformulation of $Ka_P$, which takes $p$ and $Pr$ into account:

\begin{equation}
Ka_x = \left(\frac{u^\prime}{s_0}\right)^{3 / 2}\left(\frac{L}{\delta_F(p)}\right)^{-1 / 2}\left(\frac{\delta_F(p)}{\delta}\right)^{1 / 2} \frac{1}{Pr^{1 / 2}}.
\label{eqn:Ka_x}
\end{equation}
\noindent
\noindent
We also introduce a ``width-scaled Karlovitz number'' ($Ka_w$) which multiplies $Ka_x$ by an additional tunable scaling factor based on the physical flame width:

\begin{equation}
 Ka_w = \left(\frac{t_F}{t_\eta}\right) \left(\frac{\delta_F(p)}{\delta}\right)^\theta = \left(\frac{u^\prime}{s_0}\right)^{3 / 2}\left(\frac{L}{\delta_F(p)}\right)^{-1 / 2}\left(\frac{\delta_F(p)}{\delta}\right)^{\theta+ 1/2} \frac{1}{Pr^{1 / 2}},
 \label{eqn:Ka_w}
\end{equation}
\noindent
\noindent
where $\theta$ is the scaling exponent.

Finally, we consider two Karlovitz numbers based on comparing the turbulent energy transfer rate ($\varepsilon$) with a similar quantity assembled from flame properties, $\varepsilon_F = s_0^3/\delta_F$. The first of these numbers (which we will call the ``energy Karlovitz number") was suggested by \citet{aspden2019} as a way to avoid explicitly including the Kolmogorov scale in the definition of the Karlovitz number since the smallest eddies are diluted by expansion across the flame front and don't have the ability to disrupt the flame much on their own. It is

\begin{equation}
 Ka_{\varepsilon}= \left(\frac{\varepsilon}{\varepsilon_F}\right)^{1/2} = \left(\frac{u^\prime}{s_0}\right)^{3/2} \left(\frac{L}{\delta_F}\right)^{-1/2}.
 \label{eqn:Ka_energy}
\end{equation}
\noindent
We also will consider an extension of $Ka_\varepsilon$ which explicitly accounts for the length of the turbulent cascade within the flame front by multiplying the Karlovitz number by a tunable scaling factor to model the fact that thicker flames enclose more of the turbulent cascade (at a given $Re$) and can be mixed by eddies with a larger range of sizes. We call this the ``cascade Karlovitz number":

\begin{equation}
 Ka_c = \left(\frac{\varepsilon}{\varepsilon_F(p)}\right)^{1/2} \left(\frac{\delta_F(p)}{\eta}\right)^\phi = \left(\frac{u^\prime}{s_0}\right)^{3/2} \left(\frac{L}{\delta_F(p)}\right)^{-1/2}  \left(\frac{\delta_F(p)}{\eta}\right)^\phi,
  \label{eqn:Ka_c}
\end{equation}
\noindent
where $\phi$ is the scaling exponent.

\medskip

\noindent
We will use these numbers in Section \ref{section:turbulence} to determine whether the transition from thin flames to thick flames is due to self-generated turbulence.

\section{Results}
\label{section:results}

In this section, we will consider four important questions that address the way in which the flame responds to its own self-generated turbulence. Can RT unstable flames be thickened by their own self-generated turbulence? If so, how does the flame structure change during the transition from thin to thick? Do RT unstable flames follow the turbulent combustion regime diagram? Finally, can RT unstable flames transition to distribute burning?

\subsection{Can RT unstable flames be thickened by their own self-generated turbulence?}
\label{section:turbulence}

We begin by asking whether RT unstable flames can be thickened by their own self-generated turbulence. In this subsection, we use measurements of the flame speed and flame width to show that flames transition from thin+fast to thick+slow at low $p$ and $Pr$. We argue that this transition is primarily due to self-generated turbulence because there is a clear transition in behavior with the Karlovitz number.

In order to compare the flame widths of RT unstable flames to their laminar counterparts, we must define and measure the flame width. We measure the flame width between the $T=0.1$ and $0.9$ temperature contours by iteratively breaking the flame into smaller slices, as far as the resolution will allow. We measure the isovolume of each slice and the surface areas of the two bounding temperature contours. Dividing by the surface areas gives a minimum and maximum value for the isowidth of the slice. Adding the isowidths of all slices between $T=0.1$ and $0.9$ gives a minimum and maximum value for the flame width between these contours. This method for measuring the flame width was originally developed by \citet{poludnenko2010}. Our variant is described in \citet{hicks2015}. We define the normalized flame width as the ratio of the measured and the laminar flame widths: $\delta_N = \delta^{0.9}_{0.1} / \delta^{0.9}_{0.1}(p)$.

\begin{figure}
\captionsetup{singlelinecheck = false, justification=justified}
  \centerline{\includegraphics[width=5.5in]{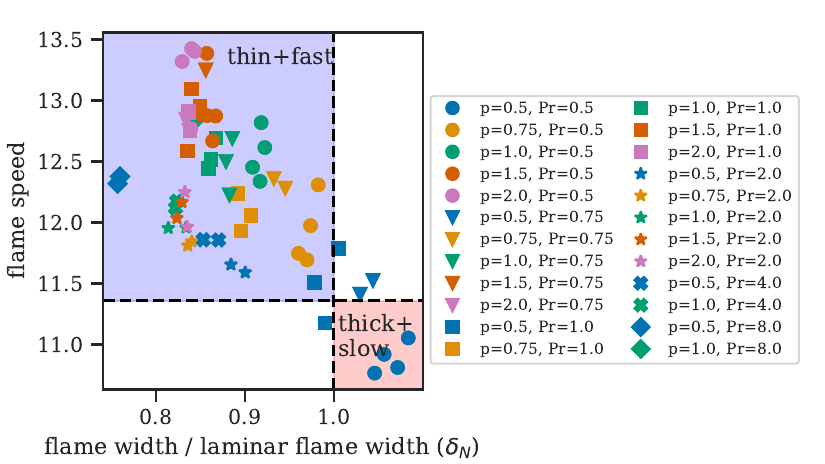}}
\caption{The time-averaged flame speed vs. the time-averaged normalized flame width for all of the simulations. Each point is a separate simulation. Points with the same color have the same $p$. Points with the same shape have the same $Pr$. Multiple simulations with different resolutions are shown for each $(p, Pr)$ parameter combination. Flames are classified as thin (thick) if their normalized flame width is less than (greater than) 1. Flames are classified as fast (slow) if their flame speed is greater than (less than) predicted by the RT flame speed model. This figure shows a transition from thin+fast flames to thick+slow flames.}
\label{figure:speedvswidth}
\end{figure}

Using the measurement of the time-averaged flame width and the measured time-averaged flame speed (as described in \citet{hicks2015}, Section 4), we can classify the simulations into two groups as shown in figure \ref{figure:speedvswidth}. First, there are the thin+fast flames. These flames propagate faster than the RT flame speed model~\citep{khokhlov1995,khokhlov1996,gamezo2003,zhang2007} predicts and they are thinner than laminar. This behavior is typical for RT unstable flames, and was first described and explored in our previous work~\citep{hicks2015,hicks2019}. More surprising are the thick+slow flames, located in the bottom right-hand side of figure \ref{figure:speedvswidth}. These flames travel more slowly than the RT flame speed model predicts and they are thicker than laminar. Overall, figure \ref{figure:speedvswidth} shows that there is a transition from thin+fast flames to thick+slow flames.

\begin{figure}
\captionsetup{singlelinecheck = false, justification=justified}
  \centerline{\includegraphics[width=5.333in]{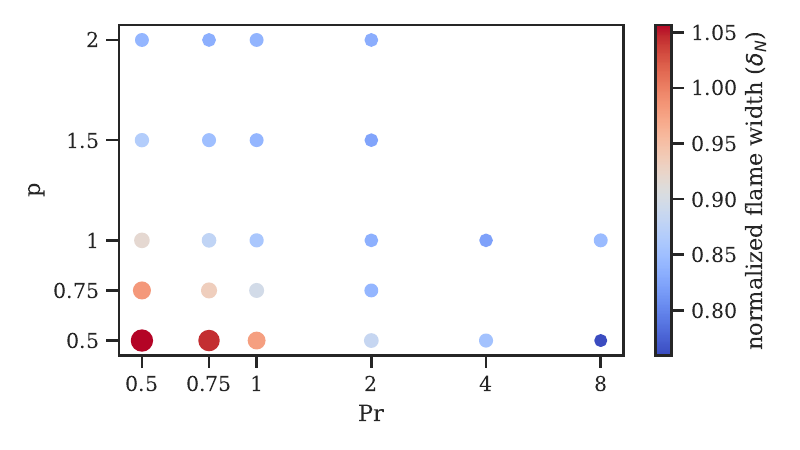}}
\caption{This figure shows that the transition from thin (bluer, smaller dots) to thick (redder, larger dots) occurs going towards low $p$ and $Pr$.}
\label{figure:parameterstransition}
\end{figure}

Where in parameter space does the transition occur? Figure \ref{figure:parameterstransition} shows the normalized flame width as a function of $p$ and $Pr$. Thicker flames have larger, redder dots. The transition from thin to thick flames occurs going towards low $Pr$ and low $p$. The thickest flames have $p=0.5$ and $Pr=0.5$.

Is the transition from thin flames to thick flames due to self-generated turbulence? If so, there should be a clear transition with one of the non-dimensional parameters that describes the strength of the turbulence. These parameters are defined in Section \ref{section:numbers}. In this analysis, we will consider the physical model underlying any given parameter to be plausible if the normalized flame width data collapse and show a clear transition from thin to thick if the parameter is increased.

\begin{figure}
\captionsetup{singlelinecheck = false, justification=justified}
  \centerline{\includegraphics[width=5.333in]{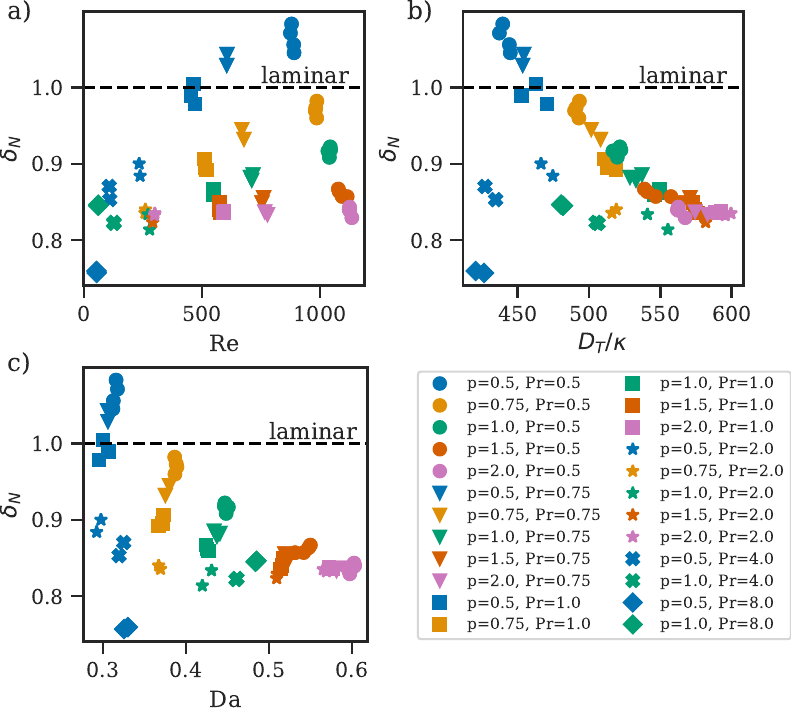}}
\caption{This figure shows how the normalized flame width depends on $Re$ (panel a), the diffusivity ratio $D_T/\kappa$ (panel b), and $Da$ (panel c). None of these parameters collapse the flame width data well.}
\label{figure:Re_diffusivity_Da}
\end{figure}

We begin with the Reynolds number and the diffusivity comparison because these numbers have been shown to control the transition from corrugated flamelets to broadened preheat/thin reaction zones in experimental studies~\citep{skiba2018}. Our data are shown in figure \ref{figure:Re_diffusivity_Da}. Although the flame width increases with $Re$ on average, there is a clear spread in the data due to $p$ (i.e. a dependence on the laminar flame width) that is not accounted for. The diffusivity ratio does a better job collapsing the data, but there is a dependence on $Pr$. In addition, the normalized flame width decreases with the diffusivity ratio instead of increasing as in turbulent combustion experiments. Likewise, figure \ref{figure:Re_diffusivity_Da}c shows a clear spread due to $Pr$ when $\delta_N$ is plotted against $Da$. Overall, figure \ref{figure:Re_diffusivity_Da} shows that the transition from thin to thick isn't well described by $Re$, the diffusivity ratio, or $Da$.

\begin{figure}
\captionsetup{singlelinecheck = false, justification=justified}
  \centerline{\includegraphics[width=5.333in]{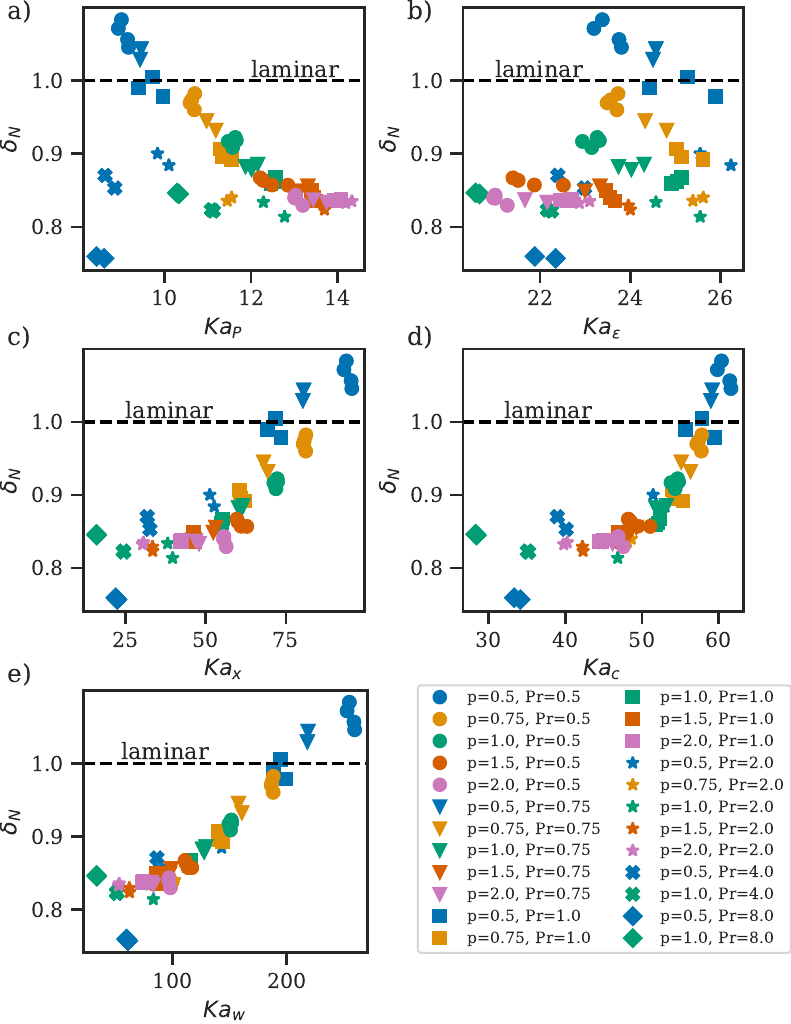}}
\caption{This figure shows how the normalized flame width depends on the Karlovitz numbers defined in section \ref{section:numbers}. The two classical definitions, $Ka_P$ (panel a) and $Ka_{\varepsilon}$ (panel b), do not collapse the data. On the other hand, new definitions that account for $p$ and $Pr$ do collapse the data well and show a thickening transition. The good collapse of the data with $Ka_x$, $Ka_c$ and $Ka_w$ suggests that the primary cause of thickening is self-generated turbulence.}
\label{figure:Ka}
\end{figure}

Next we consider whether the transition from thin to thick is well described by any of the Karlovitz numbers described in Section \ref{section:numbers}. Figure \ref{figure:Ka}a shows that the normalized flame width decreases with the classical Karlovitz number ($Ka_P$) described by \citet{peters2000}, and that there is a dependence on $Pr$. (Note that although the plots using the diffusivity ratio and $Ka_P$ look almost identical, they are actually slightly different and  $Ka_P \neq D_T/\kappa$.) Therefore, the machinery of theoretical turbulent combustion isn't adequate to describe our data.

In Section \ref{section:numbers}, we pointed out that the derivation of $Ka_P$ assumes that $Pr=1$ and only accounts for the dimensional flame width, and so we extended Peters' derivation to account for $p$ and $Pr$, deriving $Ka_x$. Figure \ref{figure:Ka}c shows that the flame with data do collapse well with $Ka_x$ (although there is clearly a small dependence on $p$ at large $Ka_x$). The plot shows that when $Ka_x$ is smaller (less than $60$ or so), the flame width remains constant at about $85\%$ of laminar. The flame width begins to thicken at around $Ka_x\sim60$. Finally, the flame become thicker than laminar at around $Ka_x\sim80$. The fact that the data collapse well with $Ka_x$ and that there is a clear transition suggest that the basic picture developed by~\citet{peters2000}, that small turbulent eddies thicken the flame, applies to RT unstable flames. However, $p$ and $Pr$ must be accounted for to describe the thickening transition.

In Section \ref{section:numbers}, we expanded on the hypothesis that turbulent eddies thicken the flame and introduced the ``width-scaled Karlovitz number" ($Ka_w$) which includes an additional tunable scaling factor based on the physical flame width. Figure \ref{figure:Ka}e shows that $Ka_w(\theta=1/2)$ collapses the data well and shows a clear transition from thin to thick flames.

Turning to Karlovitz numbers based on $\varepsilon/\varepsilon_F$, Figure \ref{figure:Ka}b shows the normalized flame width data plotted against the energy Karlovitz number, $Ka_\varepsilon$~\citep{aspden2019}. The data do not collapse well, showing that the thickening transition depends on more than just the energy transfer rate in the inertial subrange. However, if we explicitly account for the length of the turbulent cascade within the flame front by adding a tunable scaling factor to construct $Ka_c$, the data collapse well, as shown in figure \ref{figure:Ka}d for $\phi=1/4$.

Altogether, we have shown that the existing Karlovitz formulations, $Ka_P$ and $Ka_{\varepsilon}$ do not collapse our normalized flame width data well. However, if we extend the~\citet{peters2000} definition to account for $p$ and $Pr$ and derive $Ka_x$, the data collapse fairly nicely. Moreover, if we include additional factors to account for the fact that thicker flames include more of the turbulent cascade then the data collapse improves, as shown by the figures for $Ka_c$ and $Ka_w$. The precise form of the tunable factor doesn't seem too important, as long as it includes $\delta_F(p)$. In summary, our Karlovitz number study suggests that RT unstable flames are thickened by turbulent eddies, but that $p$ and $Pr$ must be accounted for to describe the transition. Physically, it seems that $\varepsilon/\varepsilon_F$ and the length of the turbulent cascade within the physical flame front are both important.

In this section, we have shown that RT unstable flames transition from thin+fast to thick+slower at low $p$ and $Pr$. This transition can be well described using an appropriately formulated Karlovitz number that accounts for both the physical laminar flame width ($\delta_F(p)$) and the $Pr$. This suggests that the flame is thickened by its own self-generated turbulence.

\subsection{How does the flame structure change during the transition from thin to thick?}
\label{section:structure}

\begin{figure}
\captionsetup{singlelinecheck = false, justification=justified}
  \centerline{\includegraphics[width=5.5in]{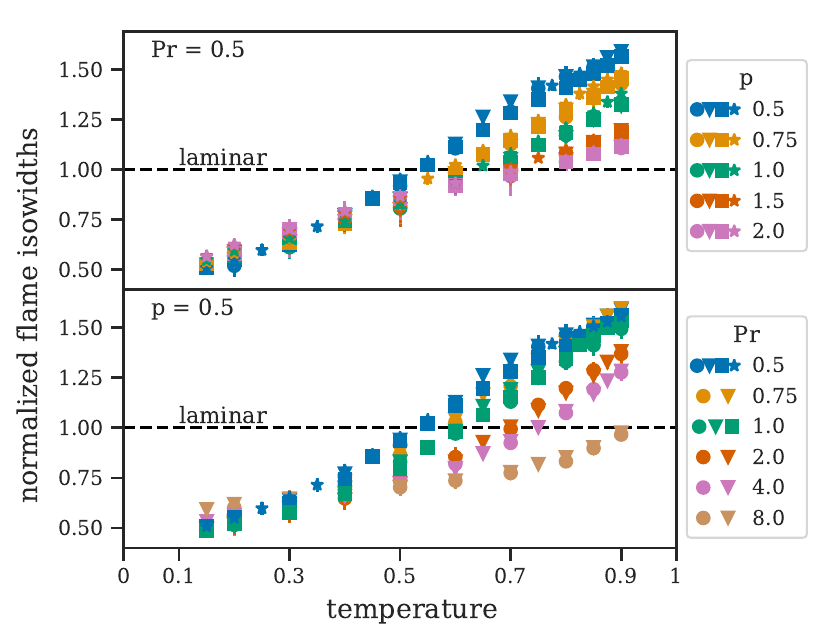}}
\caption{This figure shows how the transition from thin flames to thick flames changes the time-averaged internal flame structure. The top panel shows the transition fixing $Pr=0.5$ and decreasing $p$.  The bottom panel shows the transition fixing $p=0.5$ and decreasing $Pr$. Both plots show the widths of small slices of the flame, normalized by the widths of the same slices in the corresponding laminar $p$-flames. Values $<1$ ($>1$) indicate that the slice is thinner (thicker) than laminar. The bounding temperature values of the slices vary with resolution, so some temperature values have more data points than others. The normalized isowidth is plotted at the maximum temperature for each slice. The error bars (generally very small) show the range in the average of the normalized flame widths from the flame width calculation algorithm.}
\label{figure:flamestructure}
\end{figure}

\begin{figure}
\captionsetup{singlelinecheck = false, justification=justified}
  \centerline{\includegraphics[width=5.5in]{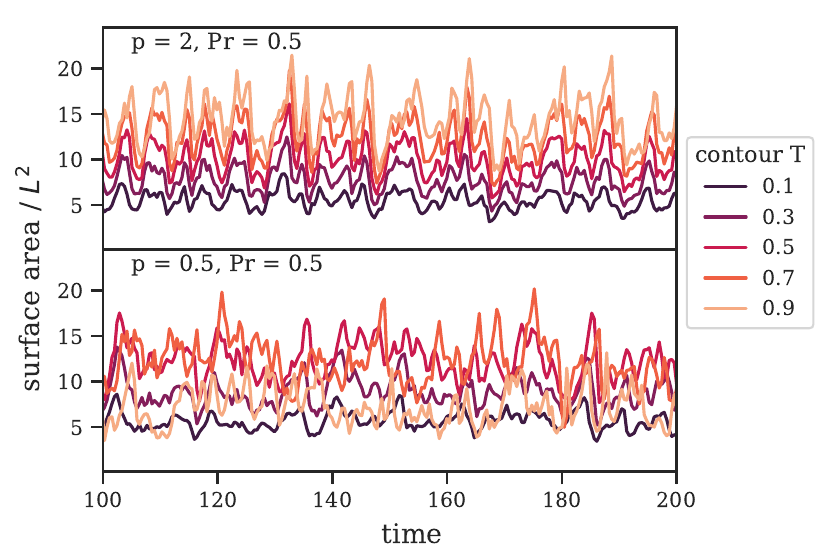}}
\caption{This figure compares the time series of the surface areas of the temperature contours within two flames: a thin flame (top panel) and a thick flame (bottom panel). Contour evolution is well-correlated in the thin flame; the flame evolves as a single unit stretched by the RT instability. In contrast, the back contours $(T=0.7,0.9)$ of the thick flame evolve independently of the front contours $(T=0.1,0.3,0.5)$ because of the self-generated turbulence that thickens the back of the flame.}
\label{figure:sa-timeseries}
\end{figure}

Next, we will investigate how and why the internal flame structure of the flame changes during the transition from thin to thick.

We probe the flame structure by dividing the flame into small slices, with the boundaries of each slice defined by a lower and upper temperature contour. To compare the flame structure to that of a laminar flame, we divide the width of each small slice by the width of the corresponding slice of the laminar flame with the same $p$-value. In addition, we use measurements of the surface areas of the bounding temperature contours to assess whether the flame is evolving as a single coherent unit.

First, we look at the transition from thin to thick by holding $Pr=0.5$ and decreasing $p$. The normalized flame isowidths are shown in the top panel of figure \ref{figure:flamestructure}. The fuel (front) and ash (back) sides of the flame behave differently. The front of the flame is much thinner than laminar, and the degree of thinning does not depend strongly on $p$ at this $Pr$. This thinning of the front of the flame is due to the strong stretching of the flame front by the RT instability. On the other hand, the back of the flame is thicker than laminar and there is a strong dependence on $p$. The thickening of the back of the flame is caused by self-generated turbulence. Thicker laminar flames (lower $p$) are more thickened by self-generated turbulence, while thinner flames are less thickened.

Next, we look at the transition from thin to thick by holding $p=0.5$ and decreasing $Pr$. The bottom panel of figure \ref{figure:flamestructure} shows that the result is qualitatively similar to holding $Pr$ fixed and decreasing $p$. The front of the flame is thinned by the RT instability. The back of the flame is thickened by turbulence, with the amount of thickening clearly increasing going from the $Pr=8$ case (which has a low $Re$ and little turbulence) to $Pr=1$. Thickening also appears to increase slightly going from $Pr=1$ to $Pr=0.5$, but the effect is smaller, either because $Pr$ is being decreased by a relatively small amount or because the flame structure might be converging as $Pr \rightarrow 0$.

The conclusion that the backs of these flames are thickened by self-generated turbulence is supported by the flame isocontour surface area time series data. Figure \ref{figure:sa-timeseries} shows the surface areas of isocontours for two flames: one thin ($p=2, Pr=0.5$) and one thick ($p=0.5, Pr=0.5$). As the thin flame evolves, the surface area curves are well-correlated and evolve together. The flame evolves as one piece that is stretched and deformed by the RT instability. The thick flame behaves differently. The front contours still evolve together, but the back contours are out of sync, both with the front contours and with each other. This is because self-generated turbulence disrupts the back of the flame.

Overall, RT unstable flames have an unusual internal flame structure: thin in front and thick in the back. This unusual internal flame structure is due to competition between the RT instability (which thins the flame) and self-generated turbulence (which thickens the flame). As the flame thickens, the temperature contours begin to desynchronize starting from the back of the flame.  This behavior is different from turbulent flames, which thicken from the front.

\subsection{Do RT unstable flames follow the turbulent combustion regime diagram?}

\begin{figure}
\captionsetup{singlelinecheck = false, justification=justified}
  \centerline{\includegraphics[width=5.5in]{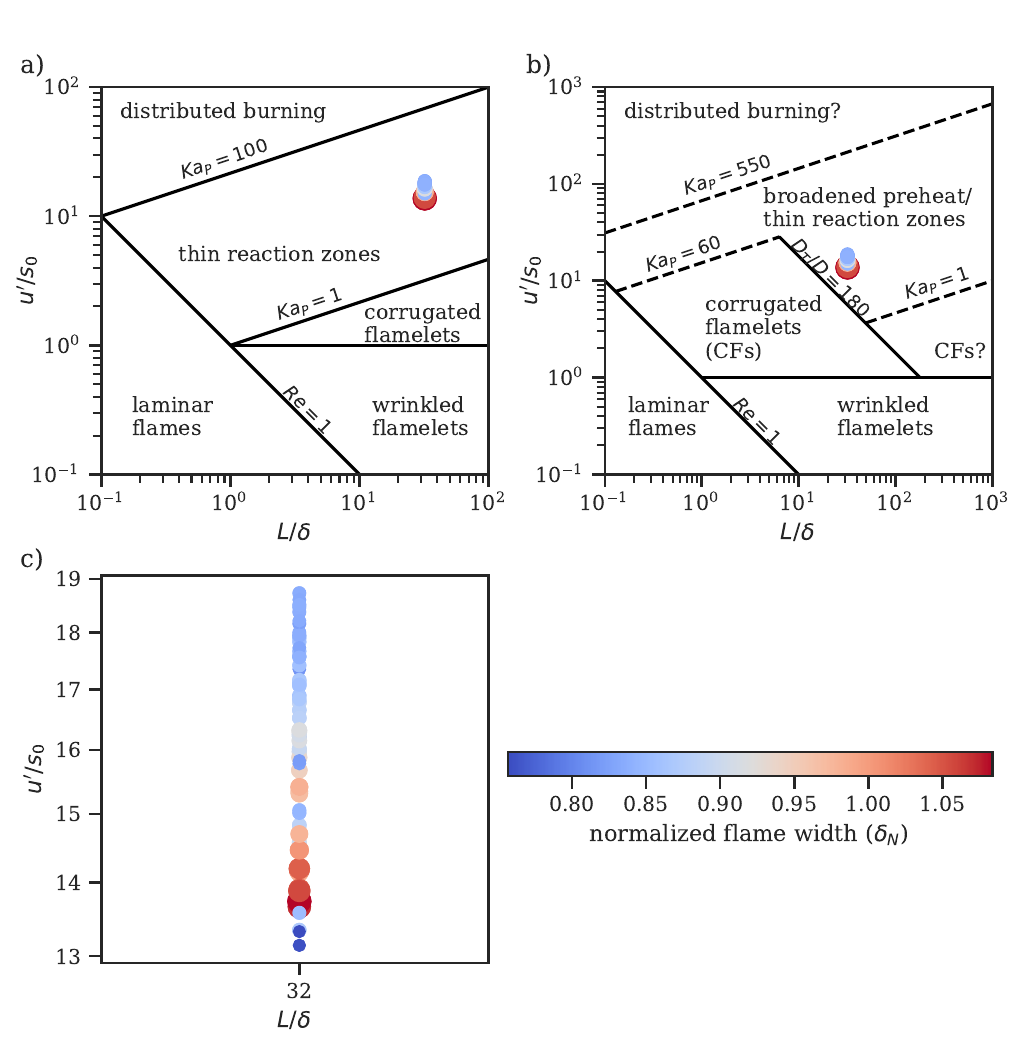}}
\caption{This figure shows our simulations placed on the traditional (panel a) and experimental (panel b) turbulent combustion regime diagrams. Panel c shows the diagrams zoomed-in to show detail. Thinner flames are represented by smaller, bluer points. Thicker flames are represented by larger, redder points. Although the simulations fall into the ``thin reaction zones'' regime on the diagrams, this regime is not a good description of their flame structure. The lack of a clear trend in the flame width and the fact the diagram does not include $Pr$ also suggest that a new regime diagram is needed for RT unstable flames.}
\label{figure:dataondiagrams}
\end{figure}

\begin{figure}
\captionsetup{singlelinecheck = false, justification=justified}
  \centerline{\includegraphics[width=5.5in]{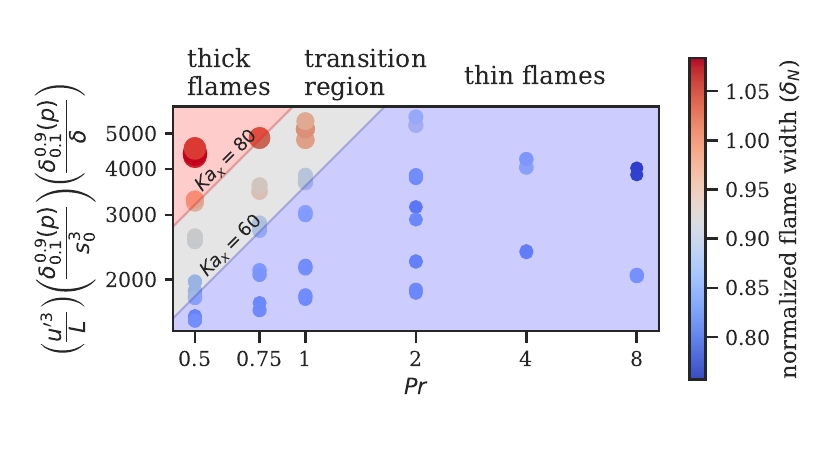}}
\caption{This figure shows our simulations placed on a new regime diagram for RT unstable flames. The diagram includes three regimes: thin, transitional, and thick flames. Where a flame falls in the diagram depends on its $Ka_x$.}
\label{figure:newdiagram}
\end{figure}

Turbulent combustion regime diagrams are useful because they classify flame behavior using simple combinations of parameters and measurements. Transitions between regimes take place at boundaries given by simple criteria involving non-dimensional parameters; for example, $Ka_P=1$ is the boundary between corrugated flamelets and thin reaction zones in the theoretical turbulent combustion regime diagram \citep{peters2000}. In this section, we will show that the theoretical and the experimentally measured turbulent combustion regime diagrams don't correctly classify the behavior of RT unstable flames with varying $p$ and $Pr$. We will propose a new regime diagram, based on our definition of $Ka_x$ (equation \ref{eqn:Ka_x}), which incorporates the effects of $p$ and $Pr$.

We begin by showing that the theoretical and the experimentally measured turbulent combustion regime diagrams don't adequately capture the behavior of RT unstable flames. Our simulations are shown on both regime diagrams in figure \ref{figure:dataondiagrams}. The simulations are grouped together closely and fall into the ``thin reaction zones'' regime on both diagrams. This reveals several issues. First, the Prandtl number, which we have shown affects the flame structure, is not accounted for in these diagrams. This is because the traditional turbulent regime diagram was derived with the assumption that $Pr=1$. The second issue is that RT unstable flames are not in the thin reaction zones regime. Thin reaction zones are flames with thickened preheat regions, but undisturbed reaction layers. In contrast, our RT unstable flames have thin fronts, but thickened backs (as we showed in section \ref{section:structure}). In addition, the RT unstable flames transition from thin-on-average to thick-on-average within this small region of the traditional regime diagram. Finally, flames thicken upwards and to the left on the traditional diagram, that is towards higher $Ka_P$. On the experimental diagram, flames thicken upwards and towards the right, that is towards higher $D_T/D$. RT unstable flames don't follow either trend, as shown in figure \ref{figure:dataondiagrams}c. Clearly, the theoretical and the experimentally measured turbulent combustion regime diagrams don't classify the behavior of RT unstable flames well and a new diagram is needed.

To develop a new regime diagram, we must satisfy certain criteria. The diagram must account for $Pr$ and $p$. There should be regions with distinct behaviors and the boundaries between these regions should be based on a sensible non-dimensional parameter. We propose a new regime diagram based on our derived expression for $Ka_x$ (equation \ref{eqn:Ka_x}) since $Ka_x$ is a direct extension of $Ka_P$ that accounts for $p$ and $Pr$ and it is a good descriptor of the transition between thin and thick flames, as shown in Section \ref{section:turbulence}. This diagram is shown in figure \ref{figure:newdiagram}. The two axes are $Pr$ and $\left(\frac{{u^\prime}^3}{L}\right) \left(\frac{\delta^{0.9}_{0.1}(p)}{s_0^3}\right) \left(\frac{\delta^{0.9}_{0.1}(p)}{\delta}\right)$, which can be interpreted as the rate at which energy is passed down the turbulent cascade divided by a similar quantity based on the laminar flame properties, multiplied by a scaling factor which describes how much thicker the physical laminar flame is than the nondimensional laminar flame width. The parameter $p$ is accounted for through $\delta^{0.9}_{0.1}(p)$. The choice of these axes comes from the expression for $Ka_x$ and the desire for lines of constant $Ka_x$ to be straight lines on this log-log plot. Two of these lines are shown in figure \ref{figure:newdiagram}; they have slopes of $1$. The lines divide three regions with different behaviors. Flames with $Ka_x \lesssim 60$ are thin on average. Flames in the transitional region ($60 \lesssim Ka_x \lesssim 80$) are beginning to thicken. Flames with $Ka_x \gtrsim 80$ are thicker than laminar on average. On the figure, the transition from thin to thick flames takes place going from lower right to upper left. This new diagram satisfies the criteria for a useful regime diagram and could be used to predict the behavior of future simulations. Similar diagrams can be constructed using $Ka_c$ and $Ka_w$, the other Karlovitz numbers that collapse the normalized flame width data. These diagrams are qualitatively similar to figure \ref{figure:newdiagram}.

In this subsection, we showed that the theoretical and the experimentally measured turbulent combustion regime diagrams don't capture the behavior of RT unstable flames. We developed a new diagram based on our derived $Ka_x$ that explicitly accounts for $Pr$ and $p$. This diagram uses $Ka_x$ to divide the flame's behavior into thin, transitioning, and thick regimes. The diagram can be used to predict whether the flame is thin or thick without explicitly measuring the flame structure.

\subsection{Can RT unstable flames transition to distribute burning?}
\label{section:Q4}

\begin{figure}
\captionsetup{singlelinecheck = false, justification=justified}
  \centerline{\includegraphics[width=5in]{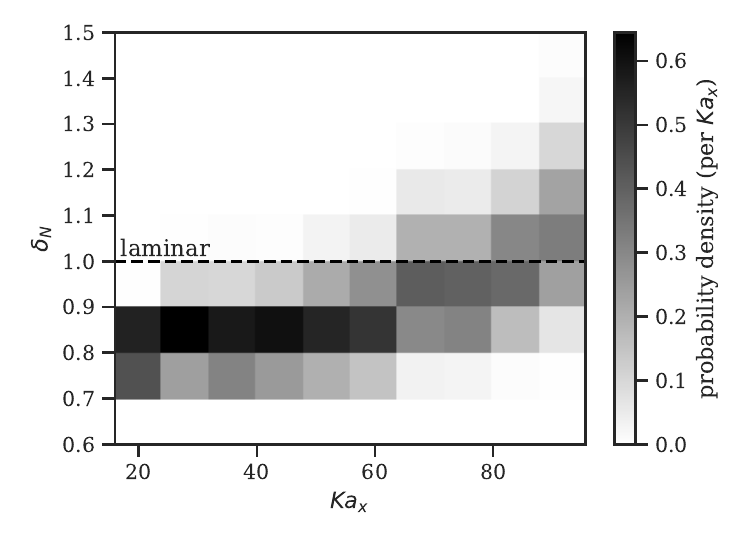}}
\caption{This figure shows how the distribution of the normalized flame width changes with $Ka_x$.  Shaded blocks show the fraction of counts that fall into a given normalized flame width range for a given range of $Ka_x$. The sum of fractions in each vertical block column adds up to $1$. At lower $Ka_x$, the distribution of flame width values is narrow and the flame is thinner than laminar. At higher $Ka_x$, the distribution widens. However, the maximum flame width increases faster than the minimum flame width. Even as the distribution widens and spreads upwards, the flame still spends time thinner than laminar.}
\label{figure:histogram}
\end{figure}

Finally, we ask whether RT unstable flames can transition to distributed burning. We will discuss this question considering two commonly used definitions of distributed burning. While our data don't definitively resolve this question, we hope that this discussion will elucidate the physical principles involved and suggest questions and approaches for future work.

\citet{aspden2011} defines distributed burning as a state where turbulent diffusion, instead of molecular diffusion, propagates the flame. Since the RT instability dominates our flow, we will consider whether or not turbulence can overcome the RT instability to become the main mechanism driving the flame upwards. We argue that this is unlikely because the RT instability is necessary to produce and sustain the self-generated turbulence.

As the flame evolves, its width oscillates with time as the flame goes through cycles of thinning and thickening. When the flame is thinner, it produces stronger turbulence because baroclinic vorticity production goes as $G\left[(\partial T / \partial x)\hat{z} - (\partial T / \partial z)\hat{x}\right]$. This strong turbulence thickens the flame, which reduces the horizontal temperature gradients and weakens the production of turbulence. As the turbulence weakens, the combination of the RT instability and burning thins out the flame again and the cycle repeats. The key point is that turbulence quenches its own production by thickening the flame.

The cyclical pattern of thickening and thinning is evident in our data, as shown in Figure \ref{figure:histogram}. This figure shows how the distribution of flame width time series values changes with $Ka_x$. As the flame begins to thicken above $Ka_x\sim60$, the distribution grows wider because the maximum flame width increases more with $Ka_x$ than the minimum flame width does. The lingering existence of thinned flame states at higher $Ka_x$ shows that the flame must repeatedly return to a thinner state to generate the turbulence that thickens the flame. This implies that a transition to a turbulence-controlled state of distributed burning is unlikely because the RT instability is required to maintain the turbulence.

An alternative definition of distributed burning is simply a flame that is much thicker than laminar because of turbulence. The precise amount of thickening isn't agreed upon, but \citet{driscoll2020} suggests that flames with thicknesses of more than 25 times laminar as qualify as distributed burning. Our data are very far from these values, with a maximum measured average value of $\delta_N=1.08$ and a highest measured maximum value of $\delta_N=1.49$. However, our finding that self-generated turbulence thickens the back of the flame suggests that thicker flames could be possible at higher $Ka_x$. Turning up $G$ would produce stronger turbulence and it seems possible that a high-$G$ flame might consist of a thin front layer and a very thick turbulent brush. This would be an extreme version of our thin+thick flame structure. Such a flame might behave like a turbulent flame with $Da < 1$, since the eddies would be produced at small scales. Behind the thin front layer, these flames may resemble distributed burning, defined as a very thickened flame.

In this subsection, we argued that whether or not RT unstable flames enter the distributed burning regime depends on the definition of `distributed burning.' A transition to a state where turbulence dominates over the RT instability seems unlikely because the RT instability is necessary to maintain the turbulence. However, our data suggest that thin+thick flames with wide, turbulently thickened flame brushes may be possible at high $Ka_x$ and $G$.

\medskip
In this section, we've considered how the interactions between the RT instability, self-generated turbulence and burning affect the flame. We've shown that RT unstable flames can be thickened by their own self-generated turbulence and that flames transition from thin+fast to thick+slow as $p$ and $Pr$ decrease. This transition happens at a critical value of $Ka_x$, suggesting that it is due to turbulence. During the transition, turbulence thickens the back of the flame, but the front of the flame remains thinned by the RT instability. Finally, we showed that a transition to turbulence-controlled distributed burning is unlikely because the RT instability, which thins the flame, is necessary to generate the turbulence that thickens the flame. However, flames with wide, turbulently thickened flame brushes may be possible.

\section{Conclusions}
\label{section:conclusions}

In this paper we investigated whether the flame structure of RT unstable flames can be described using ideas from turbulent combustion. In particular, we sought to determine whether RT unstable flames are affected by their own self-generated turbulence in the same way as laminar flames are affected by upstream turbulence. We focused on answering four specific questions: Are RT unstable flames thickened by self-generated turbulence? How does the internal structure of the flame change during the transition from thin to thick? Do RT unstable flames follow the turbulent combustion regime diagram? Can RT unstable flames transition to distributed burning? To answer these questions, we carried out a large parameter study of Boussinesq model flames. These simulations were DNS or nearly DNS in order to capture the full turbulent cascade. We varied two parameters: the Prandtl number and the physical laminar flame width. To make it possible to vary the flame width, we developed a new family of model reactions ($p$-flames) with the same laminar flame speed but different physical laminar flame widths. We measured the flame speed, the flame width and made detailed measurements of the internal structure of the flame. To make comparisons with turbulent combustion theory, we reformulated the Reynolds number, Damköhler number and Karlovitz number to account for $p$ and $Pr$. Our results show that RT unstable flames are thickened by their own turbulence at low $p$ and $Pr$, but that the structure of these thickened flames differs from turbulent flames and a new regime diagram is necessary.

The first question that we considered is whether RT unstable flames can be thickened by their own self-generated turbulence. We found a transition from thin+fast flames to thick+slow flames at low $p$ and $Pr$. This is the first time that thick+slow flames have been reported in the RT unstable flames literature and the first time that the effect of the reaction model and $Pr$ on the flame speed and flame width have been systematically evaluated. $p$ and $Pr$ both affect the flame speed and flame width. Next, we considered the cause of the thickening and whether it could be due to self-generated turbulence. We looked for a clear transition from thin to thick with various non-dimensional parameters and considered the physical model underlying any given parameter to be plausible if it collapsed the normalized flame width data. The data are not collapsed by $Re$ or the diffusivity ratio, unlike experimental turbulent combustion data~\citep{skiba2018}, suggesting that RT unstable flames are different than turbulent flames. The data are also not collapsed by $Da$, implying that the thickening transition is not driven by large scale turbulence. The data are not collapsed by $Ka_P$, so the basic machinery of theoretical turbulent combustion does not describe the transition. However, reformulating $Ka_P$ to account for $p$ and $Pr$ results in $Ka_x$, which does collapse the data. Moreover, adding additional tunable factors which model an increased influence of the physical flame width and length of the turbulent cascade within the flame improve the data collapse. Overall, these results imply that RT unstable flames can be thickened by turbulent eddies, but that $p$ and $Pr$ must be accounted for to describe the thickening transition. This implies that both $\varepsilon$ and the length of the turbulent cascade inside the flame front influence the flame thickness. In summary, RT unstable flames are thickened by their own self-generated turbulence if the physical laminar flame is thick and $Pr$ is low.

The second question that we considered is how the flame structure changes during the transition from thin to thick. We measured the flame structure by slicing the flame into thin isovolumes and then normalizing the width of each slice by its counterpart in a laminar flame. We considered two scenarios: fixing $Pr=0.5$ and varying $p$ and fixing $p=0.5$ and varying $Pr$. We found that RT unstable flames have a very unusual internal flame structure: thin in front and thick in the back. This flame structure is due to a competition between the RT instability (which thins the flame) and self-generated turbulence (which thickens the flame). This picture is supported by the time evolution of the surface area contours within the flame. As the flame thickens, the temperature contours desynchronize, starting from the back of the flame. Overall, we find that RT unstable flames and turbulent flames are both thickened by turbulence, but that the location of the thickening is different. Turbulent flames are thickened from the front while RT unstable flames are thickened from the back.

The third question that we addressed is whether or not RT unstable flames follow the turbulent combustion regime diagram. We considered both the theoretical regime diagram and the more recently developed experimental regime diagram. We showed that RT unstable flames aren't well described by these diagrams for several reasons. First, these diagrams do not include $Pr$ and we've shown that $Pr$ affects the structure of RT unstable flames. Second, the thin+thick structure of RT unstable flames is different from any of the flame structures described by the regime diagrams. It is most similar to `thin reaction zones', but thickening takes place in the back, instead of the front, layers of the flame and the front of the flame is significantly thinner than laminar. Third, there is no clear direction of thickening when we plot our data on the regime diagram and so there is no match to either the theoretical or experimental prediction. Since RT unstable flames are not described by the existing turbulent combustion regime diagrams, we developed a new regime diagram based on $Ka_x$. This diagram accounts for $p$ and $Pr$ and includes three behavioral regimes: thin flames, transitional flames and thick flames. The new regime diagram can be used to predict whether a flame is thin, transitioning or thick using the laminar flame structure, $Pr$ and a measurement of $u^{\prime}$. In summary, RT unstable flames do not follow the turbulent combustion regime diagram so we developed a new one.

Finally, we considered whether RT unstable flames can transition to distributed burning. Our data don't definitively resolve this interesting question, but they do shed some light on the physical principles involved. We argue that turbulence can never become the main mechanism for flame propagation because the RT instability is necessary to generate the turbulence. The evolution of the flame width with time is cyclical: the RT instability thins the flame and generates turbulence which quenches its own production by thickening the flame. This picture is supported by our data which show that the width distribution of the flame grows wider with $Ka_x$. So, flames must still return to a thinner state which effectively generates turbulence even as they become thicker on average. Although turbulence can't become the main mechanism for flame propagation, it is possible that thin+thick flames with very thin front layers and very thick turbulent brushes may exist at high $Ka_x$ and $G$. These flames would resemble distributed burning, as defined as a very thick flame. Overall, we find that whether RT unstable flames can transition to something resembling distributed burning depends on the definition of `distributed burning'.

So, does the structure of RT unstable flames respond to self-generated turbulence like a turbulent flame? Although there are some big picture similarities between the systems, there are important differences. Both turbulent flames and RT unstable flames can be thickened by turbulence, but the transition from thin+fast flames to thick+slow flames only happens at low $p$ and $Pr$ for RT unstable flames. Turbulent flames are thickened from the front; RT unstable flames are thickened from the back. The front of RT unstable flames remains thinned by the RT instability, even as the back layers thicken. RT unstable flames don't follow the turbulent combustion regime diagram -- $Pr$ and the physical laminar flame width (set by $p$) must be taken into account to describe their behavior. RT unstable flames probably can't transition to a completely, turbulence-controlled state of distributed burning, but may be able to develop a thick flame brush that resembles distributed burning. Overall, the structure of RT unstable flames is determined by a 3-way competition between the RT instability, turbulence, and burning. Where the RT instability wins, the flame is quite different from a turbulent flame.

\section{Discussion and Future Work}
\label{section:discussion}

What are modellers to make of these findings? In this section, we will consider the implications of our results for practical applications and describe future work.

The first important modeling question is whether it is reasonable to model RT unstable flames using flamelets. Flamelet models assume that the reacting layer can be modeled as a group of thin reaction layers embedded in a nonreacting turbulent flow field. The reaction-diffusion structure of each flamelet is described by a set of flamelet equations, which typically result from an asymptotic expansion based on the fact that the reaction layers are thin \citep{peters2000}. There are two ways to determine whether any given flame can be modelled by flamelets. The first is to check whether the reaction layer is thin and the flame propagates locally at the laminar flame speed \citep{skiba2018}. In many parts of parameter space, especially for terrestrial engineering applications, RT unstable flames probably fulfill the thinness criterion. However, the advection-diffusion-reaction structure of RT unstable flames is very different from laminar flames, a point which we will address in future work, and local flame speeds are typically not laminar. The second way to evaluate whether a reaction layer can be modelled with flamelets is to determine whether the state relations (i.e. conditional mean profiles) which result from plotting the species mass fractions against temperature (or another progress variable) are the same as the laminar state relations. \citet{driscoll2020} reports that a wide variety of turbulent flames, including flames in strong turbulence, satisfy this criterion. Since our $p$-flame model does not include a separate species equation, we can not evaluate this criterion. However, the temperature distribution of the flame is very different from the laminar temperature distribution. In summary, although RT unstable flames in most terrestrial applications are likely thin enough to qualify as flamelets (although this will need to be directly assessed for specific applications), their flame structure is so different from laminar that they may not be well described by typical flamelet equations. Assessing the validity of various flamelet equations and potentially developing new flamelet equations for RT unstable flames would be an interesting avenue for future work. For now, terrestrial applications using flamelet models should only include corrections that model turbulent thickening if $Ka_x$ is large.

A second modeling question is how simulations which directly capture the flame front should treat RT unstable flames. Our most important conclusion here is that thickened flame models have the potential to produce inaccurate results. Using an artificially thick flame model is like reducing $p$. If the other parameters are in the right regimes (especially if $Pr$ is low), the resulting flame will be too slow and overly thickened by self-generated turbulence. It is important to note that even if the thickened flame is artificially adjusted to propagate at a desired speed, the thick flame will still produce turbulent velocities that are too low. This is an issue for studies that measure the properties of the turbulence behind the flame and for studies that set the flame speed based on $u^\prime$. We believe that these conclusions are especially important because the most popular model flame is the KPP reaction, which is very thick: $\delta^{0.9}_{0.1}=18\delta$~\citep{V05}. An additional tricky point is that if $Ka_x$ is low enough, simulations with thick model reactions will produce results that are similar enough to simulations using thinner reactions that it may appear that the reaction type doesn't matter. For example, \citet{V03,V05,zhang2007,hicks2015} all reported flame speed scalings close to the predicted RT scaling at low $G$ and $L$ although they used different flame models. But this shouldn't be taken as evidence that these flame models will always produce similar results. In fact, based on our results, we expect these models to behave differently at higher $Ka_x$. Overall, the safest option for any application is simply to use an accurate reaction model instead of a thickened flame model.

Another important conclusion from this paper that affects modeling choices is that the $Pr$ affects both the flame structure and the flame speed. Many studies rely on numerical dissipation, which means that the $Pr$ can change with resolution. This can affect the behavior of the flame and make convergence harder to achieve. Many studies that set the $Pr$ simply choose $Pr=1$. This choice is relatively accurate for terrestrial flames, but very inaccurate for Type Ia flames for which $Pr=10^{-5}$. Future work on Type Ia flames should directly evaluate the effect of $Pr$ and use a realistic reaction model.

Finally, we consider how our results affect the ongoing debate about the mechanism for DDT in Type Ia supernovae. First, it is important to recognize that RT unstable flames in Type Ia burn in a much more complex environment than the simple flame-in-box simulations we've presented. In addition to the turbulence generated by the RT unstable flame, there is also pre-existing turbulence ahead of the flame. How RT unstable flames are affected by the combination of self-generated and upstream turbulence is an interesting topic for future work. In addition, burning takes place on the surface of large rising bubbles and the RT instability is unconfined. In spite of these complications, our simulations still have some potentially interesting implications for the study of Type Ia supernovae. As RT unstable flames travel towards the surface of the star, the underlying laminar flame becomes thicker and slower as the fuel density decreases. This is somewhat like decreasing $p$, which we've shown thickens the flame at low $Pr$. So, we expect the flame to become thickened by self-generated turbulence in the outer parts of the star. The flame may even transition into a distributed burning-like state, as described in section \ref{section:Q4}. This may potentially produce conditions favorable for a Zel'dovich gradient DDT. The complex flame brush could also result in a detonation due to turbulent self-acceleration~\citep{poludnenko2011,poludnenko2015,poludnenko2017_icders,poludnenko2019}. Overall, our results suggest that RT unstable flames will thicken and potentially become susceptible to multiple different detonation mechanisms in the outer part of the star. Interestingly, this implies that a detonation may be able to occur even if there is no pre-existing turbulence.

In this paper, we've shown that RT unstable flames can be thickened by self-generated turbulence, but that their flame structure is very different from turbulent (and laminar) flames. RT unstable flames don't follow the theoretical or measured turbulent regime diagrams, so studies should not infer their behavior regime or flame structure based on these diagrams. We've developed a new regime diagram which may be used for this purpose, although future work is needed to check whether there is any dependence on $G$ or $L$, which were not varied in this study. We've shown that $p$ and $Pr$ matter, so it is important for applications to simulate realistic flames and choose $Pr$ carefully. The use of thick model flames should be avoided. More work should be done to determine whether flamelet models are appropriate for RT unstable flames. Compressible studies are needed to determine whether RT unstable flames are able to detonate without pre-existing turbulence. In this paper, we described how and why the flame structure changes as the flame transitions from fast+thin to thick+slow. In our next paper, we will focus on the flame speed and explain why RT unstable flames slow down as they thicken.

%
\backsection[Acknowledgements]{Thank you to Don Lamb for suggesting that I investigate thickened RT unstable flames and for describing intriguing preliminary results from the University of Chicago FLASH Center that were never published. Thank you to R. Rosner for originally introducing me to Rayleigh-Taylor unstable flames and N. Vladimirova and A. Obabko for introducing me to the Nek5000 code and for providing the original RT unstable flames setup and scripts. I also thank R. Rosner and N. Vladimirova for interesting discussions that have influenced my thinking over the years and T. Erdmann and J. Sykes for a fascinating discussion on the applications of RT unstable flames to aviation. I am very grateful to P. Fischer, A. Obabko, and the rest of the Nek5000 team for making Nek5000 available and for giving me advice on using it. Thank you to S. Tarzia for proofreading and editing suggestions.

This work used the Extreme Science and Engineering Discovery Environment (XSEDE) \citep{xsede}, which was supported by National Science Foundation grant number ACI-1548562. This work used resources of the Advanced Cyberinfrastructure Coordination Ecosystem: Services \& Support (ACCESS) program \citep{boerner2023}, which is supported by National Science Foundation grants \#2138259, \#2138286, \#2138307, \#2137603, and \#2138296.  This work used the XSEDE and ACCESS resources Stampede2, Stampede3, and Ranch at the Texas Advanced Computing Center (TACC) through allocation PHY170024. The authors acknowledge the Texas Advanced Computing Center (TACC) at The University of Texas at Austin for providing HPC and visualization resources that have contributed to the research results reported within this paper. URL: http://www.tacc.utexas.edu. This work used the XSEDE and ACCESS resources Comet and Expanse at the San Diego Supercomputer Center (SDSC) at the University of California San Diego through allocation PHY170024 and through the Expanse Early User Program. Simulations were run using Nek5000 \citep{nek5000}. Simulation visualizations were created using VisIt \citep{visit}. Paper plots were created using matplotlib \citep{matplotlib, matplotlib-3.5.3} and seaborn \cite{seaborn}. We used Poetry \citep{poetry} for Python dependency management. Our Python analysis code used the packages pandas \citep{pandas, pandas-1.4.4}, NumPy \citep{numpy}, SciPy \citep{scipy}, and pytest \citep{pytest-7.2.0}.}

\backsection[Funding]{This work is supported in part by the M. Hildred Blewett
Fellowship of the American Physical Society, www.aps.org}

\backsection[Declaration of Interests]{The authors report no conflict of interest.}

%
\backsection[Author ORCIDs]{E.P. Hicks, https://orcid.org/0000-0001-8157-7765}
%

\appendix

\section{p-flame solution}\label{appendix:psolution}

\citet{gilding2004} considers a generalized Fisher equation of the form:

\begin{equation} \label{eqn:gilding}
T_t=\left(T^m\right)_{x x}+\left(T\left(b_0+b_1 T^p\right)\right)_x+\left\{
\begin{array}{lr}
T^{2-m}\left(1-T^p\right)\left(c_0+c_1 T^p\right) & \text { for } T>0 \\ 0 & \text { for } T=0
\end{array}\right.
\end{equation}
where $m>0,-1<p<0$ or $p>0, b_0, b_1, c_0$ and $c_1$ are real constants. Gilding shows that equation (\ref{eqn:gilding}) has a wavefront solution from $0$ to $1$ with wave speed

\begin{equation}
s_0=\theta_0+\frac{m c_0}{\theta_0}-b_0 .
\end{equation}
as long as
\begin{equation}  \label{eqn:requirements}
b_1^2+4 m c_1 /(p+1) \geq 0 \text { and } p \theta_0>0,
\end{equation}
where
\begin{equation}
\theta_0:=\frac{-b_1 \pm \sqrt{b_1^2+4 m c_1 /(p+1)}}{2} .
\end{equation}

\noindent
Furthermore, if $p>0$ and $m=1$, then the explicit wave front solution is:
\begin{equation}
T(\xi) =\left[\frac{1}{2}-\frac{1}{2} \tanh \left(\frac{p}{2} \theta_0 \xi\right)\right]^{1 / p} .
\end{equation}
where $\xi=x-s_0 t$.

Here, we use this result to construct a single-parameter family of laminar flame solutions with different flame front thicknesses but with the same laminar flame speed, $s_0=1$. We begin by noting that our ADR equation $T_t=T_{x x}+R(T)$ fits the form of equation (\ref{eqn:gilding}) with $m=1$, $b_0=b_1=0$ and $R(T)=T\left(1-T^p\right)\left(c_0+c_1 T^p\right)$. We choose $c_0=0$ so that $s_0 = \theta_0 = \sqrt{c_1 / (p+1)}$ and enforce the requirement that $s_0=1$ so that $c_1=p+1$. The reaction term then becomes

\begin{equation}
R(T)=(p+1) T^{p+1}\left(1-T^p\right)
\end{equation}

\noindent
and the explicit travelling wave solution for $p>0$ is

\begin{equation}
 T(\xi, p)=\left[\frac{1}{2}-\frac{1}{2} \tanh \left(\frac{p \xi}{2}\right)\right]^{1 / p}.
\end{equation}

\noindent
Finally, the requirements in equation (\ref{eqn:requirements}) are trivially fulfilled since $4 c_1/(p+1)=4>0$ and $p \theta_0 = p > 0$.

\section{p-flame derivatives}\label{appendix:pderivatives}

Using

\begin{equation}
\xi(T)=\frac{2}{p} \tanh ^{-1}\left(1-2 T^p\right)
\end{equation}

\noindent
it can be shown that derivatives of $T$ only depend on $T$ and $p$:

\begin{equation}
\begin{aligned}
& \frac{d T}{d \xi}=T\left(T^p-1\right) \\
& \frac{d^2 T}{d \xi^2}=(p+1) T^{2 p+1}-(p+2) T^{p+1}+T.
\end{aligned}
\end{equation}

\section{p-flames: limiting flame widths as  $p\rightarrow\infty$}\label{appendix:plimits}
\subsubsection{Thermal Flame Width}
The thermal flame width for $p$-flames is

\begin{equation}
\delta_T(p)=\left(\frac{1}{p}\right)(p+1)^{1+1 / p}.
\end{equation}
To find the flame width as $p \rightarrow \infty$, apply l'Hôpital's rule twice and use the product rule for limits and the fact that $ (p+1)^{1 / p}=\exp \left\{\ln \left[(p+1)^{1 / p}\right]\right\}$
to obtain
\begin{equation}
\lim _{p \rightarrow \infty} \delta_T(p)=1.
\end{equation}

\subsubsection{Contour Flame Width}
To find the limit for the contour flame width as $p \rightarrow \infty$, begin with
\begin{equation}
\xi(T)=\left(\frac{2}{p}\right) \tanh ^{-1}\left(1-2 T^p\right).
\end{equation}
Apply the transform $\phi=1-2 T^p$ to obtain
\begin{equation}
\xi(\phi)=\left[\frac{2 \ln T}{\ln \left(\frac{1-\phi}{2}\right)}\right] \tanh ^{-1}(\phi).
\end{equation}
Taking the limit as $p \rightarrow \infty$ is equivalent to taking the limit $\phi \rightarrow 1$ from below:
\begin{equation}
\lim _{p \rightarrow \infty} \xi(T)=\lim _{\phi \rightarrow 1^{-}} \xi(\phi)
\end{equation}
\noindent
Next, use l'Hôpital's rule to obtain
\begin{equation}
\label{equation:spacelimit}
\lim _{p \rightarrow \infty} \xi(T)=-\ln T.
\end{equation}
Therefore, the limiting shape for the flame as $p \rightarrow \infty$ is
\begin{equation}
T(\xi)= \begin{cases}e^{-\xi} & \text { for } \xi \geq 0 \\ 1 & \text { for } \xi<0.\end{cases}
\end{equation}
To obtain the limiting contour flame width, plug $T_1$ and $T_2$ into equation \ref{equation:spacelimit} and subtract:
\begin{equation}
\lim _{p \rightarrow \infty} \delta_{T_1}^{T_2}=-\ln \left(T_1\right)+\ln \left(T_2\right).
\end{equation}
Therefore, $\lim _{p \rightarrow \infty} \delta_{0.1}^{0.9}=2.1972245$...

\section{Extended Discussion: Turbulent Combustion Parameters}\label{appendix:numbers}

\subsubsection{Reynolds Number}

In this paper, we will consider two different Reynolds numbers, but reformulate the second as a diffusivity comparison. When we refer to the Reynolds number, we mean the traditional definition which in our nondimensionalization is

\begin{equation}
Re=\frac{u^\prime L}{Pr}.
\label{eqn:_appendix_Re}
\end{equation}
This is different than the definition used by \citet{peters2000} which is $Re_P=(\tilde{u^\prime} \tilde{l})/(\tilde{s}_0 \tilde{\delta}_{F,P}$), but the two definitions are the same if $Pr=1$ and $Le=1$. Note that Peters' definition for the flame width is $\tilde{\delta}_{F,P}= D/s_0$ where $D$ is the molecular diffusivity of the fluid. Thus $\tilde{\delta}_{F,P}= \tilde{\delta}$ and $\tilde{s}_0 \tilde{\delta}_{F,P} = \kappa$ when $Le=1$.  Note that our simulations do not track species, so that $D=\kappa$ and $Le=1$ in our setup. To avoid confusion, and to be consistent with the measured regime diagram in \citet{skiba2018}, we re-express the Peters Reynolds number as a comparison between diffusivities: $Re_P = D_T/D = \tilde{u^\prime} \tilde{l}/D$ where $\tilde{u^\prime} \tilde{l}$ is the turbulent diffusivity. Therefore, we will consider the ratio between the turbulent diffusivity and the thermal diffusivity

\begin{equation}
\frac{D_T}{\kappa} = \frac{\tilde{u^\prime} \tilde{l}}{\kappa} = u^\prime L.
\label{eqn:appendix_diff_ratio}
\end{equation}
Physically, this is a comparison of the two mechanisms which compete to drive the flame forwards. Note that when $Pr=1$, this is the same as $Re$, but otherwise it is different. Also, note that the \citet{peters2000} definition for $Re_P$ is only a true diffusivity comparison because $\tilde{s}_0 \tilde{\delta}_{F,P} = \kappa$, which is only true because $\tilde{\delta}_{F,P}= \tilde{\delta}$. Choosing $\tilde{\delta}_T(p)$ or $\tilde{\delta}_{0.1}^{0.9}(p)$ for $\tilde{\delta}_{F,P}$ changes the meaning of the equation. In real combustion, material properties vary from the preheat region to the ashes, so the material diffusivity should be chosen in the region that is relevant for the specific effect being studied.

\subsubsection{Damkohler Number}

The Damköhler number ($Da$) compares the turnover time of the largest turbulent eddies to the flame time, $Da = t_T / t_F$. Physically, $Da$ describes whether large scale turbulence has time to mix the flame before it burns. The flame time should depend on the physical laminar flame width; a given point in space will remain within a thicker flame longer than a thin one. Therefore, we define $t_F=\tilde{\delta}_F /\tilde{s_0}$ where $\tilde\delta_F(p)$ is a generic, physically appropriate laminar flame width. Altogether,

\begin{equation}
Da=\frac{t_T}{t_F}=\frac{s_0 L}{u^\prime \delta_F(p)}.
\label{eqn:appendix_Da}
\end{equation}
Note that the definition of $Da$ depends on $p$, but not on $Pr$. The Damköhler number can also be defined in terms of the turbulent flame width and the turbulent flame speed as in \citet{aspden2010}. We don't use that definition because the flame speed and flame width of RT unstable flames is due to both the RT instability and to turbulence.

\subsubsection{Karlovitz Number}
\label{section:appendix_Ka}

The Karlovitz number ($Ka$) describes whether small-scale turbulence has time to mix the flame before it burns. In this section, we introduce several different definitions of $Ka$. These definitions differ in the physical model of turbulence-flame interactions that they describe, and the parameters they take into account.

\subsubsection*{\textbf{Karlovitz numbers based on $t_{F} / t_\eta$}}

We begin with three Karlovitz numbers based on the physical assumption that the flame is thickened by eddies that mix the flame faster than it burns. Mathematically, these $Ka$ are based on a comparison between a flame burning time ($t_{F})$, and the turnover time of the Kolmogorov scale eddies ($t_\eta$). The first of these numbers is the classical Karlovitz number described in \citet{peters2000}, which we will call $Ka_P$. The second number ($Ka_x$) is a reformulation of $Ka_P$, which takes $p$ and $Pr$ into account. The third number ($Ka_w$) includes an additional tunable scaling factor based on the physical flame width.

\citet{peters2000} defines the Karlovitz number as the ratio between his definition of the flame time, $t_{F,P}$, and the turnover time of the Kolmogorov scale eddies, $t_\eta$:

\begin{equation}
Ka_P=\frac{t_{F,P}}{t_\eta}=\frac{\delta^2}{\eta^2}=\frac{v_\eta^2}{s_0^2}=\left(\frac{u^\prime}{s_0}\right)^{3 / 2}\left(\frac{L}{\delta}\right)^{-1 / 2}
\label{eqn:appendix_Ka_P}
\end{equation}
\noindent
This derivation assumes $t_{F,P}=\tilde{\delta}\tilde{s_0}$, where $\tilde{\delta}\tilde{s_0}= D$ and $Pr = 1$. Note that the flame width is the dimensional flame width, not the thermal flame width. $Ka_P$ can be interpreted several different ways: as a comparison between timescales, as a comparison between the Kolmogorov scale and the flame width, or as a comparison between the Kolmogorov eddy velocity and the laminar flame speed. Since all of these definitions are equivalent, it may seem unimportant to distinguish between the different physical mechanisms that they describe. However, these equivalences break down when $Pr\ne1$.

Next, we reformulate $Ka_P$ to explicitly account for $p$ and $Pr$. We follow the derivation of $Ka_P$ given in \citet{peters2000} starting with the definition $Ka = t_F / t_\eta$. Physically, we continue to assume that the flame is thickened by eddies that mix the flame faster than it burns.  We define $t_F=\tilde{\delta}_F /\tilde{s_0}$ where $\tilde\delta_F$ is a generic, physically appropriate laminar flame width. We account for $Pr$ by retaining the viscosity in the Kolmogorov turnover time, so $t_\eta=\left(\nu \tilde{l} / \tilde{u^\prime}^3\right)^{1 / 2}$. The Kolmogorov time is not only the mixing time for the smallest of the eddies, but also encapsulates information about the energy of the eddies (through the energy dissipation rate, $\varepsilon=\tilde{u^\prime}^3/\tilde{l}$) and about the total length of the turbulent cascade (which depends on $\nu$).  With these choices for $t_F$ and $t_\eta$, and using $\tilde s_0 = \sqrt{\alpha \kappa}$, we arrive at our ``extended Karlovitz number'':

\begin{equation}
Ka_x = \left(\frac{u^\prime}{s_0}\right)^{3 / 2}\left(\frac{L}{\delta_F(p)}\right)^{-1 / 2}\left(\frac{\delta_F(p)}{\delta}\right)^{1 / 2} \frac{1}{Pr^{1 / 2}}
\label{eqn:appendix_Ka_x}
\end{equation}
\noindent
where all quantities are now nondimensional.

Finally, we introduce a ``width-scaled Karlovitz number'' ($Ka_w$) which multiplies $Ka_x$ by an additional tunable scaling factor based on the physical flame width:

\begin{equation}
 Ka_w = \left(\frac{t_F}{t_\eta}\right) \left(\frac{\delta_F(p)}{\delta}\right)^\theta = \left(\frac{u^\prime}{s_0}\right)^{3 / 2}\left(\frac{L}{\delta_F(p)}\right)^{-1 / 2}\left(\frac{\delta_F(p)}{\delta}\right)^{\theta+ 1/2} \frac{1}{Pr^{1 / 2}}
 \label{eqn:appendix_Ka_w}
\end{equation}

\noindent
where $\theta$ is the scaling exponent. We showed that $\theta = 1/2$ is a good empirical choice in Section \ref{section:turbulence}. Physically, the multiplicative factor models the fact that thicker flames enclose more of the turbulent cascade (at a given $Re$). However, this factor does not explicitly account for the full range of the cascade within $\delta_F$, since it does not depend on $\eta$. Information about the Kolmogorov scale only enters $Ka_w$ through $t_\eta$.

\subsubsection*{\textbf{Karlovitz numbers based on $\varepsilon / \varepsilon_F$}}

Finally, we consider two Karlovitz numbers based on comparing the turbulent energy transfer rate ($\varepsilon$) with a similar quantity assembled from flame properties, $\varepsilon_F = s_0^3/\delta_F$.

The first of these numbers was suggested by \citet{aspden2019} as a way to avoid explicitly including the Kolmogorov scale in the definition of the Karlovitz number since the smallest eddies are diluted by expansion across the flame front and don't have the ability to disrupt the flame much on their own. Aspden's formulation, which we will call the ``energy Karlovitz number", is

\begin{equation}
 Ka_{\varepsilon}= \left(\frac{\varepsilon}{\varepsilon_F}\right)^{1/2} = \left(\frac{u^\prime}{s_0}\right)^{3/2} \left(\frac{L}{\delta_F}\right)^{-1/2}.
 \label{eqn:appendix_Ka_energy}
\end{equation}
\noindent
Note that \citet{aspden2019} uses the thermal flame width $\delta_T$ as the physical flame width but we will use $\delta_{0.1}^{0.9}(p)$. The idea behind this definition for the Karlovitz number is that the interaction between the flame and turbulence is driven by the interaction between the inertial subrange (which depends on $\varepsilon$, but not $\nu$) and the flame. Therefore $\varepsilon$ is important, but the viscosity is not. According to this definition, the exact length of the turbulent cascade and the size of the smallest eddies aren't important. Note that this definition is equal to $Ka_P$ if $s_0 \delta_T=\nu$. This is true if $Pr=1$ and if $\delta_T=\delta$, which is not typically true.

We will also consider an extension of $Ka_\varepsilon$ which explicitly accounts for the length of the turbulent cascade within the flame front. We include this effect by multiplying the Karlovitz number by a tunable scaling factor. This factor models the fact that thicker flames enclose more of the turbulent cascade (at a given $Re$) and can be mixed by eddies with a larger range of sizes:

\begin{equation}
 Ka_c = \left(\frac{\varepsilon}{\varepsilon_F(p)}\right)^{1/2} \left(\frac{\delta_F(p)}{\eta}\right)^\phi = \left(\frac{u^\prime}{s_0}\right)^{3/2} \left(\frac{L}{\delta_F(p)}\right)^{-1/2}  \left(\frac{\delta_F(p)}{\eta}\right)^\phi,
  \label{eqn:appendix_Ka_c}
\end{equation}
\noindent
where $\phi$ is the scaling exponent. We call this the ``cascade Karlovitz number". Physically, we expect $\phi < 1$ since the smallest turbulent eddies make less difference to thicker flames. We showed that $\phi = 1/4$ is a good empirical choice in Section \ref{section:turbulence}.

\bibliographystyle{jfm}

\end{document}